\documentclass[twocolumn]{article}

\usepackage{iftex}
\ifPDFTeX
  \usepackage[utf8x]{inputenc}
  \DeclareUnicodeCharacter{00A0}{ }
\fi

\usepackage{mathtools, cuted}
\usepackage{amsmath}
\usepackage{amssymb}
\usepackage{graphicx}
\usepackage{hyperref}
\usepackage[breakable]{tcolorbox}

\usepackage{titlesec}

\titleformat*{\section}{\large\bfseries}
\titleformat*{\subsection}{\large\bfseries}
\titleformat*{\subsubsection}{\large\bfseries}
\titleformat*{\paragraph}{\large\bfseries}
\titleformat*{\subparagraph}{\large\bfseries}

\usepackage{listings}
\usepackage{xcolor}

\definecolor{codegreen}{rgb}{0,0.6,0}
\definecolor{codegray}{rgb}{0.5,0.5,0.5}
\definecolor{codepurple}{rgb}{0.58,0,0.82}
\definecolor{backcolour}{rgb}{1,1,1}
\definecolor{orange}{rgb}{1, 0.65, 0}

\lstdefinestyle{pystyle}{
    backgroundcolor=\color{backcolour},   
    commentstyle=\color{codegreen},
    keywordstyle=\color{orange},
    numberstyle=\tiny\color{codegray},
    stringstyle=\color{codegreen},
    basicstyle=\ttfamily\footnotesize,
    breakatwhitespace=false,         
    breaklines=true,                 
    captionpos=b,                    
    keepspaces=true,                 
    numbers=left,                    
    numbersep=5pt,                  
    showspaces=false,                
    showstringspaces=false,
    showtabs=false,                  
    tabsize=2
}
  
\lstnewenvironment{pythonscript}[1][]
{
\pythonstyle
\lstset{\small #1}
}
{}
\providecommand{\tightlist}{%
   \setlength{\itemsep}{0pt}\setlength{\parskip}{0pt}}

\title{Raytracing Black Holes with AI Accelerators}

\author{Thomas Fischbacher\footnote{Google, Brandschenkestrasse
  110, 8002 Zürich, Switzerland -- \texttt{tfish@google.com}}, Nicholas Kessler\footnote{Department of Computer Science, ETH Z\"urich, Universit\"atstrasse 6, 8092 Z\"urich, Switzerland -- \texttt{nkessler@ethz.ch}}}
\date{}

\begin{document}

\onecolumn

\maketitle

\begin{abstract}
  We show how software infrastructure that was mainly built to
  power AI applications also can be used to great benefit for other
  purposes -- by demonstrating a fully self-contained JAX-based black
  hole raytracer in under 180+200 lines of Python code plus documentation.

  Major motivations for showcasing such a construction are (a) to show
  to professional physicists how Machine Learning (ML) accelerators
  can greatly simplify building numerical applications for GPU
  accelerator hardware with very modest coding effort and programming
  expertise -- even if the problems at hand do not involve ML per se,
  (b) to make a meaningful part of General Relativity (GR) accessible
  to computer science professionals who might not have realized that
  mathematical background they developed on Deep Learning problems may
  have brought this theory within easy reach for them, and (c) to
  inspire students and educators to explore the potential of an
  approach that represents physical law, i.e. relations between local
  differential quantities, not in terms of differentiable symbolic
  expressions, but in terms of differentiable algorithms.
\end{abstract}

\begin{figure}
\centering
\includegraphics[width=0.45\textwidth]{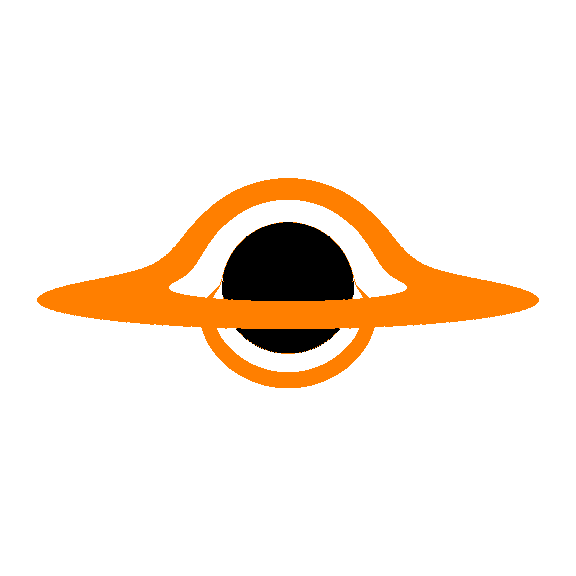}\kern2em
\includegraphics[width=0.45\textwidth]{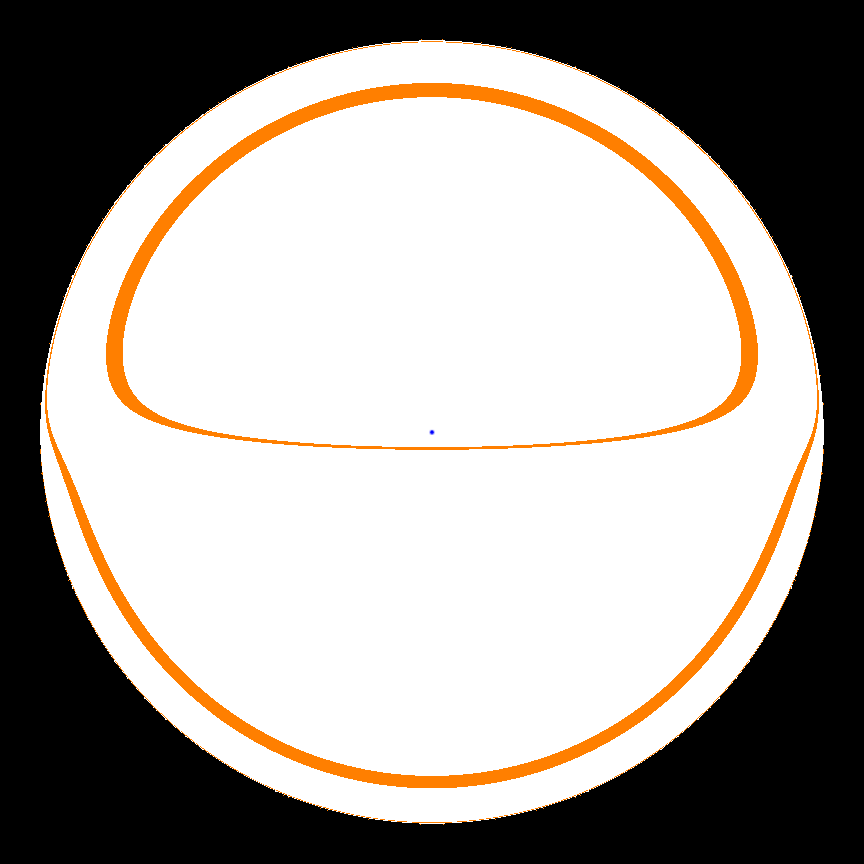}
\caption{Left: Image of an (effectively-)Schwarzschild
  (i.e. minimally-rotating) black hole with accretion disc around it,
  as seen from a stationary-distance camera about $6^\circ$ above the
  ring plane, at a coordinate distance of about 15 Schwarzschild radii
  from the coordinate center of the hole.  Light bending around the
  hole enables us to see all of the upper side of the disc and also
  part of the underside of the disc at the same time.  Right: Image
  taken from an identical second at-stationary-distance camera, at a
  coordinate distance of about 1.05 Schwarzschild radii, and looking
  `up', away from the hole (i.e. towards the first camera), same
  viewing angle. The black hole takes up most of the sky, and we only
  see the rest of the universe through a small circular window, `as if
  from deep down a well'. The small blue dot marks the center of the
  picture.  Since also here, we are above the disc, we need to `look
  down' a bit to see its upper side, but `looking up`, we can see the
  part of the disc on the other side of the hole. If we look `way
  down' from the center out of our very small window into the rest of
  the universe, we see the underside of the disc -- which closes to a
  very thin line. Very close inspection reveals further images of the
  disc even closer to the black area. For the photographer to escape
  from this situation, they would have to accelerate to
  near-light-speed and fly towards this window, every other strategy
  would make them fall into the hole.}
\label{fig:firstpage}
\end{figure}

\twocolumn

\section{Introduction}

Two important closely related recent developments have led to a
situation where advanced scientific ideas have come within reach of
understanding for much larger audiences than ever before -- where
``understanding'' here is to mean ``being able to fully understand the
construction, verify calculations, and correctly perform some own
calculations that can be verified by others''.

One of the great intellectual accomplishments of the 20th century was
the construction of a quantitative theory of gravity as -- loosely
speaking -- ``space-time elasticity'' in 1915 by
Einstein~\cite{einstein1915feldgleichungen} and, almost simultaneously,
Hilbert~\cite{hilbert1915grundlagen}. While since the inception of this
``General Theory of Relativity'', it has captivated the mind of people
all around the world, from its earliest days on, there was this great
divide between those who had an interest in the physics and its
implications but have not themselves mastered the mathematical
machinery needed to perform calculations -- and those few who have. As
a famous anecdote goes, when British physicist Arthur Eddington
received a comment in 1919 at a Royal Society meeting that he were
``one of only three men who actually understood general relativity'',
he responded ``who is the third?''~\cite{isaacson2007einstein}.
Correspondingly, in particular due
to its stark implications, General Relativity did receive a lot of
public interest throughout much of the 20th century -- with
considerable cross-fertilization through science fiction, which helped
make at least some otherwise rather obscure ideas from theoretical
physics such as about black holes common knowledge. Much of this
naturally has happened at a ``no equations'' level -- i.e. without
interested people developing a quantitative grasp of the physics.

In more recent times, other major scientific breakthroughs have led to
an explosion in our collective capacity to handle problems that
involve highly advanced mathematics. Especially in the recent past,
the wide and growing interest in topics related to Machine Learning
has led to a situation where many professionals in computer science
and adjacent disciplines have a basic repertoire at their disposal
that includes useful levels of familiarity with (multi)linear algebra
and multivariate calculus. Overall, the benefits of this side effect
of the ``AI Revolution'' still appear to be generally far
under-appreciated, and what better opportunity to make this point than
by explaining some key constructions underlying General Relativity --
at a quantitative level, where at the end we will have a
fully-understandable General Relativity raytracer for rendering (for
example) black hole geometries in a way that every reader with a
reasonable degree of proficiency in ML accelerator frameworks such as
JAX~\cite{jax2018}, PyTorch~\cite{pytorch2019}, or
TensorFlow~\cite{tensorflow2016} should find accessible\footnote{%
The JAX-based construction presented here is based on an earlier
TensorFlow-based example from course on ``Machine Learning and Machine
Learning Technology for Physicists'' that the first author taught at
the Albert Einstein Institude in Potsdam in 2022. Lecture notes
for this course are available at:
\url{https://github.com/fischbacher/IMPRS2022_ML/blob/main/IMPRS2022_ML_ALL.pdf}}?

\subsection{Everyone likes a good story}

Still, interesting as such an approach might be in its own right, it
mostly only exemplifies some deeper, and much more important insights:
First, the content of the physics curriculum can be regarded as a
collection of ``stories'' -- perhaps of a rather unusual and highly
specialized kind -- that are told from person to person, and from
generation to generation, but just as there are many different ways of
telling a story, there also are many different ways to learn possibly
very different useful things (some perhaps not at all intended by the
original author) from any given story. Second, the way we normally
express and think about the quantitative elements of such ``physics
stories'' -- the maths that we typically express with formulas which
only are accessible to a small group of people who mastered the art of
reading them -- are only auxiliary tools that we developed to talk
about nature, and nature itself does not care about the language we
use to communicate our ideas about it. Third, both ``new perspectives
on how to read equations'' and ``new ways to express quantitative
relations that add something new to the already-widely-employed
formula language'' are invaluable for our ability to narrate these
stories in ways that unlock more depth. One famous such 20th century
revolution in our ability to communicate about physics was the
introduction of ``Feynman Diagrams'' (see e.g.~\cite{schweber2020qed})
-- this gave us new story-telling devices\footnote{In the
(critical-as-often) words of Murray Gell-Mann, ``In [Quantum
  Electrodynamics], as in other quantum field theories, we can use the
little pictures invented by my colleague Richard Feynman, which are
supposed to give the illusion of understanding what is going on in
quantum field theory.''~\cite{mulvey1980nature}}.

Another new story-telling device that may well have just as
profound implications is the realization that, for some uses,
quantitative relations can occasionally be expressed better by
\emph{algorithms}~\cite{sussman2015structure} than \emph{equations},
and in particular the ability to lift calculus \emph{from terms to
algorithms}, using insights such as sensitivity backpropagation (also
known as ``reverse-mode automatic
differentiation''~\cite{speelpenning1980compiling}), is a powerful and
currently under-appreciated storytelling device.

Ostensibly, the intent of this article is about demonstrating how such
new approaches to ``physics story-telling'' that put a stronger
emphasis on algorithmic descriptions (by ``showing code'') than on
mathematical equations can be useful to make the difficult but
exciting topic of General Relativity accessible at a quantitative
level to a much larger audience than before. At a deeper level, it is
just as much about showcasing a new way of ``narrating physics'' that
hopefully will inspire others to make other very interesting but
deeply technical topics more broadly accessible. From
Bardeen-Cooper-Schrieffer theory of superconductivity to the quantum
mechanics of the hydrogen atom and also orbital dynamics of satellite
swarms or the functioning of a field effect transistor -- there are
many other gems of scientific insight that perhaps currently are not
as widely regarded as exciting topics by a broad audience as General
Relativity is, but would allow for exciting new story-telling using
techniques very similar to those employed here.

In any case, even readers with only a cursory interest in General
Relativity may well find the geometry techniques that this article
develops in compact form useful for all kinds of applications where a
non-Cartesian coordinate system can be used to great effect, but the
mathematical pre-requisites so far looked too intimidating to them.

\subsection{Structure of the Material}

Naturally, a presentation that emphasizes the use of ``language for
expressing quantitative constructions which is so precise that a
computer can handle it'' (i.e. computer code) for the purpose of
explaining physics will want to show both code snippets and formulae
in the main text. With respect to code, it is highly desirable to have
a complete, coherent, self-contained construction be part of the
presentation somewhere. Yet, it also is desirable to minimize
repetition that does not benefit the exposition.

Our approach to resolving this tension is to show skeletal key
definitions as code snippets in figures in the main text, alongside
a more elaborate, self-contained and properly-documented complete
construction in appendix~\ref{app:FullCode}. As we are
using Alphabet's JAX machine learning framework~\cite{jax2018}, code
examples will be given in Python. The ``JAX'' framework allows us to
compile computations specified in Python to fast GPU/TPU
code. Readers who want to follow the construction step-by-step can
load the Python module accompanying this article into a CPython
interpreter\footnote{This does not even require installing Python, but
can be done, for example, via a free-of-charge
\url{https://colab.research.google.com} Google Colab sandbox.} and
check and explore all definitions in the main text as they appear.

This construction aims to be presentable in its entirety using no more
than 180 lines of code plus 200 lines of documentation. Naturally,
given an educational tool like this, there are many follow-up
questions that one would want to explore since they are then within
easy reach of minor modifications. In the interest of brevity, these
are not presented in the main article, but contained as ``extras'' in
the code repository accompanying this article.

As this may be helpful to the reader, we briefly outline the structure
of our presentation. 

\begin{itemize}

\item (Section~\ref{sec:geometry}): {\bf Doing Geometry without Cartesian Coordinates}.
  General Relativity describes spacetime curvature. Curved spaces
  (for which the surface of a sphere is the simplest example) in general
  do not admit Cartesian coordinates. We develop basic machinery for working
  with generic non-Cartesian coordinate systems for use with either flat or
  non-flat space.

\item (Section~\ref{sec:parallel}): {\bf Parallel Transport in Curved
  Spaces}.  On some generic non-flat space, there is a notion of
  ``local velocity vectors of curves going through a given point
  $\mathcal{P}$''. The geometric properties of curvature manifest when
  one tries to answer questions related to the idea of ``how can we
  parallel-transport such a local velocity-vector associated with
  point $\mathcal{P}$ to some nearby point $\mathcal{Q}$ -- and then
  ultimately back to the starting point around a loop''. Intuitively,
  this corresponds to ``carrying a spear around the globe while moving
  it without rotating it''.

\item (Section~\ref{sec:computingChristoffel}): {\bf Computing
  Parallel Transport Coordinate Corrections}.
  Here, we discuss technical aspects of quantitatively obtaining
  parallel transport vector coordinate corrections, using the JAX
  framework.
  
\item (Section~\ref{sec:geodesics}): {\bf Geodesics as Autoparallel Curves}.
  Starting at some given point ${\mathcal P}$ and picking some initial
  velocity, we can ask the question: ``What happens if we keep
  parallel-transporting the velocity vector in the direction of the
  velocity-vector itself?'' This idea leads to the notion of
  ``geodesics'', which can be thought of as being ``as straight lines
  as possible on some curved space''. If the velocity-vector's speed
  is lightspeed, the corresponding geodesic describes a light ray in
  curved space-time.

\item (Section~\ref{sec:ode}): {\bf Numerical Differential Equations}.
  Here, we see how to solve the geodesic equation numerically.
  ``Given the momentary motion-state of a photon (so, position and
  direction), as well as a way to compute the rate-of-change of its
  motion-state as a function of its motion-state, how can we compute a
  numerical approximation to its path through space-time?''. This
  ``auxiliary mathematics'' section covers basics at the level of depth
  needed to complete the construction.
  
\item (Section~\ref{sec:raytracing}): {\bf Aspects of Raytracing in a
  Black Hole Geometry}. In two subsections, we first move from
  curved two-dimensional surfaces to the 3+1-dimensional curved
  space-time of General Relativity. We then sketch the structure of
  some mathematically not very deep additional bits and pieces that
  need to be fleshed out in order to obtain a working
  raytracer.

\item (Section~\ref{sec:curvature}): {\bf On Curvature}. This section
  explains some selected important aspects of Einstein's construction
  of General Relativity (i.e. ``how matter curves space-time''). Our
  immediate interest here is in obtaining a test and a debugging tool
  for our construction -- if we did everything correctly, then the
  Schwarzschild background geometry must be a ``source strength =
  zero'' vacuum solution. We also use the opportunity to add some
  detail to the conceptual differences between approaches based on
  symbolic differentiation and approaches based on
  algorithm-differentiation.

\item (Section~\ref{sec:images}): {\bf Images}. This section shows and
  discusses images produced with the raytracer.

\item (Section~\ref{sec:grphysics}): {\bf The Physics of General
  Relativity}. This section adds some general physics background on
  the key ideas underlying General Relativity, most importantly the
  equivalence between gravitating mass and inert mass.
  
\item (Section~\ref{sec:conclusion}): {\bf Conclusion and Outlook} concludes our explorations.
  
\end{itemize}

Appendices add detail that are relevant, but more technical, and
likely better studied separately, so that the main story can be
followed without much distraction. They also provide the
diligently-documented full construction.

\subsection{Related Work}

It is useful to place this educational effort in the context of other
related work. On one end of the spectrum, there are impressive
real-time GR raytracers, often built using WebGL for browser-based
accessibility.  Projects like Bruneton's~\cite{bruneton2020} WebGL2
black hole visualization achieve remarkable speed, but typically focus
on specific, highly symmetric spacetimes (Schwarzschild or Kerr) where
significant analytical simplifications are possible. On the other end,
research frameworks exist like GYOTO~\cite{vincent2011} or
RAPTOR~\cite{bronzwaer2018} that can model light paths through more
general spacetimes and include effects like magnetic fields or
radiative transfer. However, their generality often comes with
substantial complexity, making them less suitable for learning the
fundamentals from the ground up. This article aims for a different
focus: a bare-bones, minimalistic framework designed to be fully
explained and understandable while keeping the codebase under
400 lines.

\section{Doing Geometry without Cartesian Coordinates}\label{sec:geometry}

We use a familiar problem to motivate our explorations into analytical
geometry in situations where Cartesian coordinates are an unavailable
luxury: projecting the curved surface of the Earth onto flat maps. Via
pondering this task, we introduce the core ideas behind non-Cartesian
coordinates and understanding geometry in curved spaces.

Out of the many possible map-projections we could use as an example,
we pick the stereographic projection of the region near Earth's
southern pole onto the plane -- mostly since this particular
projection also has other prominent uses in mathematics and hence is
useful to know about.

\begin{figure*}[ht]
\centering
\includegraphics[width=\textwidth]{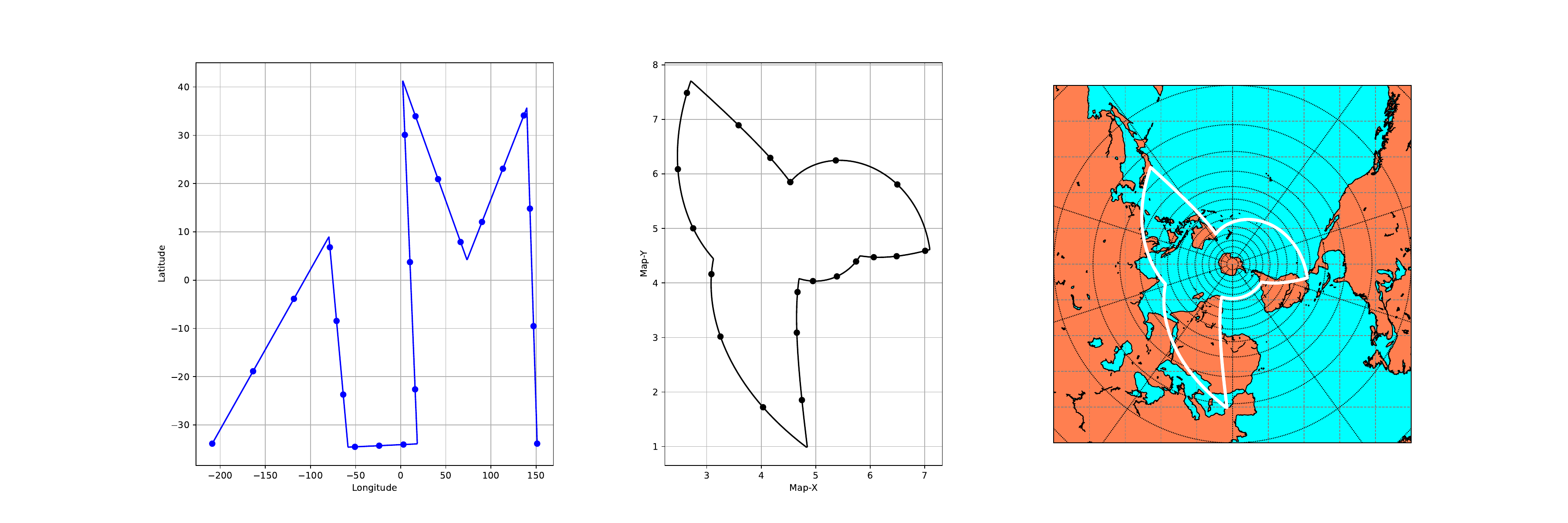}
\caption{Three views of a flight path through multiple southern-hemisphere
  cities. Left: the path as straight-line segments in latitude-longitude
  coordinates. Middle: the same path under the stereographic projection,
  where the segments now appear curved. Right: the projected path with
  continent outlines for geographic reference.}
\label{fig:map-views}
\end{figure*}

It is created by imagining Earth's surface as a (semi-transparent)
sphere of radius 1, with a light source placed at the North Pole
\((x,y,z)=(0, 0, 1)\). This light casts a shadow of every point on the
Southern Hemisphere onto a flat plane tangent to the South Pole at
\(z=-1\). This shadow forms our map. In this projection, the South
Pole is at the map's center, and the North Pole is ``at an infinite
distance''. The code in appendix~\ref{app:plotflightpaths} shows code
for rendering a path that consists of straight line segments in the
(latitude, longitude) coordinate system in terms of what they would
look like under this map. Unlike `lines of constant compass heading'
(`loxodromes', or `rhumb lines'), these lines are not particularly
relevant for navigation: they here merely serve as one ad-hoc example
for using different mathematical descriptions, or ``pictures'', for
one particular process -- motion on Earth's surface in a specific way.
Figure~\ref{fig:map-views} illustrates this idea with a flight path
through six southern-hemisphere cities, shown in three different ways:
as straight-line segments in latitude-longitude coordinates, as the
same path under the stereographic projection (where the segments now
appear curved), and with continent outlines for geographic reference.

\subsection{Motion in Curved Spaces}

Our ultimate goal is to describe the motion of light in curved
spacetime. To do this, we use a coordinate system which effectively
acts as a ``geometry-distorting map''. This closely parallels the
problem of describing an airplane's flight path on a map of the Earth.
In both scenarios, we need to relate physical properties as
experienced locally (like the momentary velocity-vector of a light ray
or the acceleration felt by a passenger) to the coordinate quantities
we deduce from the map. We therefore look at such flight paths as more
intuitive examples.

Consider an airplane at some time \(t\). It has a velocity, but what
does ``velocity'' actually mean here? In high school physics, velocity
is simple: a change in position over time, given for example as
`meters per second' changes of x/y/z-coordinates. This implicitly
makes use of a Cartesian coordinate system.

On Earth's surface, there is no universally agreed-upon
nice-to-use-for-all-purposes coordinate system. We can describe its
instantaneous velocity in at least three different ways:

\begin{enumerate}
\def\labelenumi{\arabic{enumi}.}
\item
  Latitude/Longitude Velocity: We can state how quickly the plane's
  latitude and longitude are changing. For example, its velocity might
  be (+0.5 degrees/hour, +2.0 degrees/hour). This is the rate of change
  of the coordinates themselves, what we call a `coordinate velocity'. It
  is a pair of numbers: \((d(\text{lat})/dt, d(\text{lon})/dt)\).
\item
  Map \((x,y)\) Velocity: Alternatively, we can look at the flat map
  projection and state how quickly the plane's \(x\) and \(y\)
  position on the map (for example: in centimeters!) are
  changing. This would be a different pair of numbers, \((dx/dt,
  dy/dt)\).
\item
  Physical Velocity: We can describe the velocity as a local observer on
  the ground would measure it, using a coordinate system that is
  orthonormal at that specific point. This is the physically meaningful
  velocity, expressed in intuitive units like ``meters per second East''
  and ``meters per second North''. This is the velocity we need to
  calculate real-world physical quantities like kinetic energy.
\end{enumerate}

The first two are easy to compute from our data, but the third is the
one with direct physical meaning. Though these three representations use
different numbers and units, they all describe the exact same state of
motion. This points to a single, underlying geometric object that
represents the true velocity, independent of any coordinate system. This
object is the velocity vector.

In geometry, a vector is often introduced as an arrow having a specific
length and direction. Importantly, a vector is an abstract concept: an
arrow drawn on a graph from \((0,0)\) to \((1,2)\) is considered the
same vector as one drawn from \((5,5)\) to \((6,7)\). They are
equivalent because they share the same length and direction.

We can apply this same thinking to our flight paths. Imagine a specific
point \(\mathcal{P}\) on our map, say Tahiti, a small island in the South Pacific.
Our Sydney-to-Panama flight passes through \(\mathcal{P}\) with a specific
instantaneous velocity. Now, imagine another flight path, say from Tokyo
to Buenos Aires, that also happens to pass through Tahiti. If, at the
moment it is over the island, the second plane has the exact same
velocity as the first, we consider their motion-state at that point to be
identical.

Any number of different flight paths could pass through Tahiti,
but if they all share the exact same velocity at that point, they are
all representatives of a single, abstract geometric object: a tangent
vector at \(\mathcal{P}\). This vector represents the instantaneous velocity at
that specific location, independent of the journey's origin or
destination.

The collection of all possible tangent vectors at a single point \(\mathcal{P}\)
forms what is called a tangent space, which we can denote \(T_\mathcal{P}\). For
our 2D map, this is a 2D vector space. This means we can perform
standard vector operations like addition and scalar multiplication on
these vectors. For instance, in our map's \((x, y)\) system:

\begin{equation}
\vec{v}_C = \vec{v}_A + \vec{v}_B = (v_{Ax} + v_{Bx},\; v_{Ay} + v_{By})
\end{equation}

In particular, for two airplanes passing over the same point on
Earth's surface at the same time (at different, constant altitude), we
can use vector subtraction to determine relative velocity of one plane
as seen from the other.

The remarkable property is that this operation is independent of our
chosen coordinates. The tangent space \(T_\mathcal{P}\) behaves just like the
familiar 2D plane of vectors from introductory physics. It is a flat
space of all possible ``arrow-like'' velocities that can exist at
point \(\mathcal{P}\).

With a firm grasp of tangent vectors, we can ask a physical question. If
our flight path describes an airplane of mass \(m\), what is its kinetic
energy at a time \(t\)?

\begin{equation}
E_k = \frac{1}{2} m \, |\vec{v}|^2,
\end{equation}

To answer this, we need the plane's true speed-squared, \(|\vec{v}|^2\).
Here we face a crucial problem. If we take our map-coordinate velocity,
\(\vec{v} = \left(\frac{dx}{dt}, \frac{dy}{dt}\right)\), and compute its
squared length naively using the Pythagorean theorem, we get a
meaningless result:

\begin{equation}
\left( \frac{dx}{dt} \right)^2 + \left( \frac{dy}{dt} \right)^2
\end{equation}

This would be the speed of the dot moving on our flat map, not the speed
of the actual airplane over the curved Earth. Due to the map's
distortion, we need a tool that tells us how to compute the true,
geometric squared length of a tangent vector from its coordinate
representation at any point \(\mathcal{P}\).

This tool is an important object in General Relativity. It is called the
metric tensor, often written as \(g\). For our purposes, we can think of
the metric as a machine that, for any point \(\mathcal{P}\), provides a specific
rule for calculating the squared length \(|\vec{v}|^2\) of any vector
\(\vec{v}\) in the tangent space \(T_\mathcal{P}\).

A complete description of the local geometry at a point \(\mathcal{P}\) should
include not just the lengths of vectors, but also the angles between
them. This information is encoded in the scalar product (or dot
product), \(\vec{v} \cdot \vec{w}\). At first, it might seem that we
need a new, separate rule for computing this product.

Fortunately, a mathematical relationship called the polarization
identity saves us the effort. It shows that if we know how to compute
the squared length of any vector, we can automatically compute the
scalar product between any two vectors. The identity is:

\begin{equation}
\vec{v} \cdot \vec{w} = \frac{1}{4} \left( |\vec{v} + \vec{w}|^2 - |\vec{v} - \vec{w}|^2 \right)
\end{equation}

This directly follows from \(|\vec{v} \pm \vec{w}|^2=(\vec{v} \pm \vec{w})\cdot(\vec{v} \pm \vec{w})\).
With this identity, our entire effort to understand the local geometry
of our curved space, calculating lengths, angles, and kinetic
energies, boils down to a single, focused task:

We must find the rule that gives the true physical squared length,
$ |\vec{v}|^2 $, for any tangent vector $ \vec{v} $
at any point \(\mathcal{P}\).

This rule, which connects coordinate-based vectors to physical lengths,
is precisely what the metric tensor provides.

\subsection{The ``Metric Tensor''}

We established that we need a rule to compute the true, physical
scalar product of two tangent vectors at any point \(\mathcal P\). For
two vectors
\(\vec{v}\) and \(\vec{w}\) with coordinates \((v_0, v_1)\) and
\((w_0, w_1)\), this rule will take the general form of a
weighted sum of products of coefficients\footnote{
In Physics, most texts use indices $1, 2,\ldots$,
while in computing, it is more natural to start index counting at 0.
Modern Relativity texts also start index counting at 0, with $x_0=t$
being the time-coordinate. We hence use indices that start at
0 throughout.}:

\begin{equation}
\begin{array}{lcl}
\vec{v} \cdot \vec{w}&=&M_{00} v_0 w_0 + M_{01} v_0 w_1 +\\
&&M_{10} v_1 w_0 + M_{11} v_1 w_1
\end{array}
\end{equation}

where the coefficients \(M\) can change depending on our location
\(\mathcal P\).
The symmetry of the dot product (\(\vec{v} \cdot \vec{w} =
\vec{w} \cdot \vec{v}\)) requires that \(M_{01} = M_{10}\). These
coefficients \(M_{ij}\) can be arranged into a symmetric \(2 \times 2\)
matrix. This matrix is the so-called ``metric tensor''
evaluated at the point
\(\mathcal P\). As the scalar product is linear in each of its two arguments,
we can understand $M_{ij}$ as the `geometric' scalar product of the
$i$-th coordinate-vector with the $j$-th coordinate-vector at $P$ --
also in situations where these coordinate-vectors are not
mutually-orthogonal, where we then see off-diagonal entries
$M_{ij}\neq 0$ for $i\neq j$.

It is important to understand that these coefficients are not
constants, they are themselves functions of our coordinates \((x,
y)\).  The ``metric'' is the rule or function that tells us what these
numbers are at any given point\footnote{We do note that the term
``metric'' is also used differently in mathematics. Here, ``metric''
is short hand for ``metric in (pseudo-)Riemannian Geometry''.}.

By picking a specific point \(\mathcal P\), we evaluate that function
to get a matrix that shows the complete local geometry at that one
spot. In general, when we talk about $P$ as a ``physical location'',
we may well use very different maps to describe that point -- and for
each map, $P$ will in general have different coordinates. Likewise,
the value of the metric tensor at $P$ (so, the matrix) will be
different depending on which map we use, but the ``physics
predictions'' -- such as the kinetic energy computed from the scalar
product of the velocity vector with itself -- will have to align,
irrespective of what map we are using to describe the physics.

\subsection{The ``Line Element''}\label{sec:ds2}

So, how do we find the metric \(M\) for our spherical Earth? In
geography, we use coordinates of latitude (\(\theta\)) and longitude
(\(\varphi\)). A velocity vector is then given by its coordinate
components, which are the rates of change
\(\left( \frac{d\theta}{dt}, \frac{d\varphi}{dt} \right)\). For
simplicity these rates will be denoted with a prime:
\((\theta', \varphi')\). In physics literature, one also often
sees `over-dot' notation \((\dot\theta, \dot\varphi)\) for
rates-of-change in time.

For a sphere of radius \(R_E\), the true squared speed \(|\vec{v}|^2\)
of a velocity vector \(\vec{v} = (\theta', \varphi')\) at a point with
latitude \(\theta\) is given by the formula:

\begin{equation}|\vec{v}|^2 = R_E^2 \cdot \left( (\theta')^2 + (\varphi')^2 \cos^2 \theta \right)\end{equation}

This expression is equivalent (in terms of its information content) to
what is known as the ``line element'' in physics. The crucial term
is \(\cos^2 \theta\). It correctly accounts for the fact that circles
of constant latitude shrink as we move from the equator (\(\theta =
0^{\circ}\)) towards the poles (\(\theta = \pm 90^{\circ}\)). At the
equator, \(\cos(0) = 1\) and the formula is simple. Near the
poles, \(\cos(\theta)\) approaches zero, so even a large change in
longitude \(\varphi'\) corresponds to a very small physical distance.

This single formula for the squared length, \(|\vec{v}|^2\), is all we
need. As we saw with the polarization identity, from this one function,
we can derive the full scalar product.

Let us illustrate this with an example: Consider a boat traveling
northeast at a constant speed of one knot (one nautical mile per hour,
the nautical mile being 1/60 of 1/360 of earth's circumference).

If the boat is at the equator (\(\theta = 0^\circ\)): Traveling
northeast for one minute means the boat moves approximately 21 meters
east and 21 meters north. This movement corresponds to a change of
roughly one arc-second in both latitude and longitude, as the distances
for one minute of arc are the same in both directions.

If the boat is at 60° southern latitude (\(\theta = -60^\circ\)): The
circle of constant latitude is now half the circumference of the
equator, since \(\cos(-60^\circ) = 0.5\). If the same boat again
travels 21 meters east and 21 meters north, the latitude still changes
by one arc-second (since the north-south distance for a minute of arc is
constant everywhere). However, to cover that same 21 meters eastward,
the longitude now must change by two arc-seconds, not one.

This example shows why the \(\cos^2\) term is needed. The line element
formula correctly translates the non-uniform coordinate grid into the
true squared speed, which is what we need for physical calculations.

The code in figure~\ref{fig:jax_sphere} shows how we can express this
line element for the sphere as a simple JAX function and then use JAX
to obtain objects such as an algorithmic representation of the metric
tensor coordinates as a function of position\footnote{In many
production-grade-code situations, we either can get away with not
explicitly transforming JAX-supporting callables with
\texttt{@jax.jit}, leaving these details entirely to JAX, or are
well-advised to think more deeply about cache-state management by
making stateful callables be owned by class instances.}
Operationally, there are different ways how to think about the
relation between this `line element' and the local scalar product: If
we wanted to have a callable function which, given an array of
coordinates of a point in space plus two arrays containing the
coordinates of two velocity-vectors at that point (with coordinates
referring to the local coordinate system), the most straightforward
way to compute the scalar product is likely to utilize the
polarization identity. The code illustrates how to do this in generic
form by computing a \emph{lexical closure} -- a nested function that
retains information about the context in which it is defined (here:
the line-element function). JAX fully supports computations that
utilize such closures. While we could then obtain the metric tensor by
repeatedly calling this function once per pair of basis vectors of the
local coordinate system, it is more straightforward to directly obtain
a function that computes the metric tensor's matrix entries as a
function of coordinates by applying JAX's automatic differentiation to
the JAX-function computing the length-squared. Given the position --
here, $(\theta, \phi)$, we compute (half of) the matrix of all
possible 2nd derivatives (the ``hessian'') w.r.t. the rate-of-change
variables $(\theta', \phi')$, which we conveniently evaluate at the
$\theta'=\phi'=0$ origin\footnote{Readers should be able to at least
verify numerically on examples that the scalar product defined in
terms of these matrix entries indeed computes the length-squared of
arbitrary vectors at the given point of choice.}: due to the scalar
product being bi-linear in these rate-parameters, the matrix of 2nd
derivatives is independent of $(\theta', \phi')$.

It is noteworthy that we are here capturing physical meaning not
exclusively in the form of symbolic expressions which we then
manipulate with multivariate calculus. Rather, we are capturing
physical meaning in the form of executable code, and then use
transformations that involve obtaining rates-of-change via algorithmic
differentiation. As explained earlier, the purpose of our presentation
is to demonstrate the power of this nowadays generally under-utilized
approach.

\begin{figure*}[t]
\begin{lstlisting}[language=Python]
import jax
from jax import numpy as jnp

R_E = 6371e3  # [m], Earth's radius.
rad_from_deg = jnp.pi / 180

# The specific geometry of interest:
@jax.jit
def jax_L2_geography(coords_lat_lon, rates):
  lat, lon = jnp.asarray(coords_lat_lon, dtype=jnp.float64)
  v_theta, v_phi = (jnp.asarray(rates, dtype=jnp.float64)
                    * rad_from_deg)
  r_ring = R_E * jnp.cos(lat * rad_from_deg)
  return R_E**2 * v_theta**2 + r_ring**2 * v_phi**2

# Some generic helpers for obtaining algorithmic descriptions of
# geometry from other such algorithmic descriptions.

def scalar_product_func_from_L2_func(L2_func):
  def sp_func(coords, v1, v2):
    v1a = jnp.asarray(v1); v2a = jnp.asarray(v2)
    return 0.25 * (L2_func(coords, v1a + v2a) -
                   L2_func(coords, v1a - v2a))
  return sp_func

def metric_func_from_L2_func(L2_func):
  d2_L2_by_dxi_dxj_func = jax.hessian(L2_func, argnums=1)
  def metric_func(coords_pos):
    return 0.5 * d2_L2_by_dxi_dxj_func(coords_pos, jnp.zeros_like(coords_pos))
  return metric_func

# {pos_coords}, {v1_coords}, {v2_coords} -> {scalar product}
jax_sprod_geography = jax.jit(
  scalar_product_func_from_L2_func(jax_L2_geography))

# Applying the generic tools:

jax_metric_geography = jax.jit(
  metric_func_from_L2_func(jax_L2_geography))

# >>> print(jax_metric_geography(jnp.array((60.0, 25.0))))
# [[1.23643117e+10 0.00000000e+00]
#  [0.00000000e+00 3.09107793e+09]]
\end{lstlisting}
\caption{Python JAX code for latitude/longitude line element on earth's surface}
\label{fig:jax_sphere}
\end{figure*}

There are different ways to compute this hessian using different
combinations of forward- and reverse-mode differentiation for each of
the differentiation steps, with interesting performance
trade-offs\footnote{Details can be found in the github code repository
accompanying this work}. Overall, the problem at hand only involves
dependency on very few (coordinate-)parameters. Frameworks like JAX
provide major benefit especially when used on problems where the
number of parameters is well above 100 or so.

\subsection{Coordinate Transformations and Scalar Products}

When computing the numerical entries of the coefficient-matrix
for the scalar product \(M\), which gives us 
``geometrically meaningful local scalar product between
(velocity-)vectors in non-Cartesian coordinates as
\(\vec v\cdot\vec w = \vec v^{\,T} M \vec w = \sum_{i,j} v_i M_{ij} w_j\),
we find fairly large entries. This is
readily understood as a consequence of asking for a meters-squared
answer when the input is changes to latitude and longitude:
one degree of latitude change corresponds to
\(1/360\) of Earth's circumference (about \(40\,000\,{\rm km}\)), so
order-of-magnitude \(100\,{\rm km}\). A latitude-change of one degree
per second would hence mean a contribution of
approximately\footnote{Actually about 10\% more on each factor,
so we are here under-estimating the product by about 20\%,
entirely in alignment with the actual computed numbers.}
\((100\,{\rm km}/{\rm s})\times (100\,{\rm km}/{\rm s})=10^{10}\,{\rm m^2}/{\rm s^2}\),
explaining the order-of-magnitude of the corresponding matrix entry
shown at the end of the code in Figure~\ref{fig:jax_sphere}.  At
$60^\circ$ northern latitude, each degree of longitude change amounts
to half the distance that it would at the equator, so the scalar
product of a velocity-vector that changes longitude and not latitude
at a given rate of degrees per second with itself will be
$(1/2)\cdot(1/2)=1/4$ of that of a corresponding vector that changes
latitude and not longitude at the same degrees-per-second rate. Again,
this is exactly what we see in the ratio of matrix entries in the
example shown in figure~\ref{fig:jax_sphere}.

Also, we only get non-zero entries on the diagonal.
This is so since locally, latitudes are perpendicular to longitudes,
but depending on where we are on the planet, one degree of
longitude-change (at same latitude) can amount to a much smaller
distance than one degree of latitude-change (at same longitude).

If we have a map, and know the mapping of \((x,y)\) image-position
map-coordinates to latitude and longitude, this also allows us to
translate an \((x,y)\) position rate-of-change to a latitude/longitude
rate-of-change, and find the matrix that allows us to compute the local
velocity-vector scalar product for vector coordinates that describe
\((x,y)\)-coordinate rates-of-changes. In general
(such as when using a map rotated by an unusual angle), we
find a non-diagonal symmetric matrix \(M\).

A worked out example for converting Cartesian coordinates \((x,y)\) to polar
coordinates \((r, \theta)\) can be found in appendix~\ref{app:polar}.
  
\section{Parallel Transport in Curved Spaces}\label{sec:parallel}

Our first ingredient was the metric, which lets us measure lengths and
angles locally. Our second theoretical tool is the concept of
parallel transport, which will allow us to define ``straight lines'' in
a curved space.

On a flat sheet of paper, a straight line is simple. A key property is
that its tangent vector always points in the same direction. If we walk
in a rectangle and come back to our starting point, a vector we
carried with us will point in the exact same direction. But what does
``same direction'' mean on a curved surface like a sphere?

Imagine we stand on the equator and hold a spear pointing eastward,
along the line of the equator. Now, we take it on a trip around a
closed triangular loop, being careful to never locally ``turn'' the
spear:

\begin{enumerate}
\def\labelenumi{\arabic{enumi}.}
\tightlist
\item
  First, we walk north along a line of longitude until we reach the
  north pole.
\item
  Next, from the north pole, we walk south along a different line of
  longitude -- for this example, we pick the direction of the spear, so
  90 degrees to the right from the prior marching direction.
\item
  Finally, when we hit the equator again, the spear points due south.
  We walk west along it to return to our starting point. If we do so
  while not turning the spear, it will always point south.
\end{enumerate}

Our spear started by pointing east, but at the end of this procedure,
it points south. Even though at every step, we moved without turning it.
The trip through curved space induced a net 90-degree rotation.

This phenomenon, a vector changing its orientation after being
parallel-transported around a closed loop, is called holonomy. It is a
consequence of the space being curved. Loosely speaking, the amount of
rotation is proportional to the curvature enclosed by the loop. On a
flat sheet of paper, we never get an actual rotation out of parallel
transport along closed curves, the holonomy is ``trivial''.

This shows that the idea of ``keeping a vector's direction constant''
on a curved surface is tricky. Parallel transport is the precise
mathematical procedure for doing this. A geodesic is then a path that
parallel-transports its own tangent (velocity w.r.t. the path's
curve-parameter) vector, it is ``as straight as possible'' in a
coordinate-independent sense.

If a path looks straight in some particular coordinate system (like
latitude/longitude), this does not necessarily mean it is a
geodesic. For example, on a sphere, most ``constant latitude'' paths
are not geodesics. Great circles are geodesics, and most of them
involve changes to both latitude and longitude.

While the example above only motivated the idea of geodesics in terms
of a plausible example, it turns out that one can introduce this as a
mathematical concept even by exclusively referring to local
measurements that an inhabitant of -- possibly curved -- space can
perform. It makes sense to discriminate between ``intrinsic'' and
``extrinsic'' geometric properties -- such as curvature -- of some
curved space:

\begin{itemize}
\tightlist
\item
  ``Extrinsic'' properties make reference to how the space in question
  is embedded into some higher dimensional space. A cylinder mantle is
  a 2d space that is bent in 3d space.
\item
  ``Intrinsic'' properties can be defined without making reference to
  any such embedding.
\end{itemize}

To understand this, consider an observer in 3d space versus a 2d
creature living on a surface. The 3d observer could take a poster and
glue it perfectly onto a cylinder without stretching or tearing
it. Any geometric construction performed by inhabitants living on the
2d world of the poster would survive the ``bending'': the sum of
angles in any triangle is still 180 degrees, the length of the
boundary of the shape made up by all points less than 5 cm from a
given point still is $2\pi\cdot5\,{\rm cm}$, etc. To an inhabitant of
the poster on the cylinder who cannot reach out into the 3rd
dimension, their world still is geometrically indistinguishable from a
flat plane.

The sphere is fundamentally different. An external observer cannot
flatten a piece of a sphere's surface without distortion. The 2d
creature on the sphere can discover this intrinsic curvature on their
own. If they start at a pole and walk a set distance \(r>0\) in all
directions to mark out a circle, they can then measure its
circumference, \(C\). They will find that \(C\) is less than the
\(2\pi r\) they would expect on a flat plane.

To be precise, if the sphere has radius \(R\) (here a 3d
concept), the circumference the 2d creature measures is given by the
formula $C=2\pi R\sin(r/R)$. Since here the sine function will always
be less than its argument (i.e., \(0<\sin(x)<x\) for $0<x<\pi/2$),
it follows that \(C=2\pi R\sin(r/R) < 2\pi R(r/R)=2\pi r\).
This discrepancy is how they can discover their world is intrinsically curved.
Crucially, for a very small circle (where \(r\) is tiny relative to $R$), the
approximation \(\sin(r/R)\sim r/R\) becomes very accurate, so \(C\sim
2\pi r\). This shows that any curved space looks flat when only
performing small-scale experiments that are not sensitive enough
to resolve discrepancies at the curvature's own length scale
(which here is the sphere's radius).

This matters because we are the creatures living inside our \(3+1\)
dimensional spacetime. We cannot ``step outside'' to see how it is bent.
All of our physics must be described intrinsically, using only
measurements we can make from within. This is why we need a definition
of parallel transport that relies solely on the intrinsic geometry of
spacetime.

So, how do we describe parallel transport mathematically?

Let us say we take a vector \(\vec{v}\) and transport it for a small
distance `without stretching or rotating' (we will make this fully
precise later) along a path \(\gamma\) which at the given
path-coordinate $s_*$ has a
position-coordinate-rate-of-change-(w.r.t.-path-coordinate) vector
\(\vec{c}\). When we do this, the components of \(\vec{v}\) in our
coordinate system (like (lat, lon) or \((x, y)\)) will change. This is
because our coordinate grid lines are themselves bending and
stretching.

We want to find the rate of change of \(\vec{v}\)'s components,
\(dv_i/ds\), as we move along the path \(\gamma\). This rate of change
must depend on two things: the vector \(\vec{v}\) we are carrying, and
the velocity vector \(\vec{c}\) of the path we are moving along.

It turns out that this rate of change depends linearly on the vector
\(\vec{v}\) being transported (If we transport a vector that is twice
as long, its components will change twice as much, and transport also
must respect vector addition: ``The sum of two transported vectors
must equal the transported sum''). This gives us:

\begin{equation}\frac{dv_i}{ds} = \sum_j T_{ij} v_j\end{equation}

where \(T\) is some matrix that depends on the point on the curve
currently in focus.

Furthermore, the change is also linear in the transport velocity
vector \(\vec{c}\) (If we move along the path twice as fast, the rate
of change of \(\vec{v}\)'s components should double -- and also, ``if,
at the given point, we know the rates-of-change
(w.r.t. curve-parameter) for a set of paths with unit velocity along
each of the coordinate directions, we can compute the rate-of-change
for a path with arbitrary velocity-vector''). This means the matrix
\(T\) must itself be a linear function of the coordinates in \(\vec{c}\).

We can write this relationship out explicitly for each entry
\(T_{ij}\)\hspace{0pt} as:

\begin{equation}T_{ij}(\vec{c}) = \sum_k \tilde C_{ijk} c_k\end{equation}

The numbers \(\tilde C_{ijk}\) are the correction-coefficients that
connect the velocity vector \(\vec{c}\) to the ``coordinate-adjustment
matrix'' \(T\). These \(\tilde C\)-coefficients depend only on our
position in the space, not on \(\vec{v}\) or \(\vec{c}\), and allow us
to perform small-distance parallel transports of any vector along
curves that go through the current point in any direction. They
capture all the information about how the coordinate system itself is
``bending and turning'' at that point.

Putting it all together, we get the full expression for how the
components of a vector change under parallel transport:

\begin{equation}\left( \frac{d v_i}{d s} \right) = \sum_j T_{ij}(\vec{c}) \, v_j = \sum_{j,k} \tilde C_{ijk} \, v_j \, c_k\end{equation}

The most widely used convention is to not go with the equation above
as a definition of the $\tilde C_{ijk}$, but instead express this
relation in terms of their negatives $C_{ijk}:=-\tilde C_{ijk}$. The
deeper reason is that when we talk about derivatives (which one does a
lot in General Relativity), we generally do this by pulling a
``displaced'' object (such as: a vector) backwards(!) towards a chosen
reference point, and then compare the backwards-displaced object with
the object at the reference point.  Due to this built-in ``reversal of
the displacement'', the more frequently used formula for ``coordinate
system corrected'' derivatives of position-dependent vectors looks
more straightforward when expressed in terms of $C_{ijk}$ rather than
$\tilde C_{ijk}$.

These vector coordinate correction-coefficients \(C_{ijk}\) provide a
notion of parallel transport along a curve for vectors, and, through
compatibility requirements, then also for any other geometric object
that can be defined in terms of vectors -- they define the ``affine
connection''. In General Relativity, these coefficients are the famous
Christoffel symbols (of the second kind)\footnote{In usual general
relativity notation, these are the \(\Gamma^i{}_{jk}\).}. They are the
fundamental objects that define parallel transport and, as we will
see, give us the geodesic equation.\footnote{A note on notation: A
reader consulting other sources might find the last index on the
Christoffel symbols to link to the vector being transported, not the
vector along which to transport. However, the connection in General
Relativity is symmetric \(C_{ijk}=C_{ikj}\), and this symmetry ensures
the final sum is the same regardless of the index order.}

To make this less abstract, let us return to the very problem we
encountered in the previous section: the boat traveling northeast on
the Earth's surface. We established that the coordinate grid itself is
not uniform. At $60^\circ$ southern latitude, an eastward boat trip
covering a fixed small distance resulted in twice the change in
longitude coordinate as it did at the equator at $0^\circ$
latitude. This showed a problem: the coordinate values are not a
reliable guide to true distance or direction.

Parallel transport is the tool that solves this problem. It provides the
rules for how a vector's components must change as we move to counteract
the stretching of our coordinate system. This procedure allows us to
keep the vector pointing in the ``same direction'' in a physically
meaningful way.

Let us revisit a point $P$ at $60^\circ$ latitude and consider our two
local basis vectors: a northward vector (\(\vec{e}_{\theta}\)) and an
eastward vector (\(\vec{e}_\phi\)). A key insight is that in our
coordinate system for $N$-dimensional space, we can imagine at every
point $N$ `coordinate lines' running through that point along which
(at least locally) all coordinates except one are constant, and the
varying coordinate can be taken as the curve-parameter. There then is
a 1:1 correspondence between the local rates-of-change along these
coordinate-curves and the local basis vectors. Now, we will analyze
what happens when we parallel-transport our two basis vectors along a
path moving north.

Transporting the northward vector (\(\vec{e}_\theta\)): As we move
north along a line of longitude, the ``north'' direction aligns with our
path. The vector is carried along without any rotation relative to the
path. Therefore, its components in our coordinate system do not change.
The correction is zero.

Transporting the eastward vector (\(\vec{e}_\phi\)): Here, the failure
of the coordinate system to be Cartesian becomes apparent. As we move
north, the lines of longitude converge.  This means that a fixed
physical distance (e.g., 1 kilometer) to the east corresponds to a
progressively larger change in the longitude coordinate, \(\phi\). To
keep a physical eastward vector constant (parallel transport), its
component in the \(\phi\) direction must therefore increase to
compensate for this geometric effect: $\tilde
C_{\phi,\theta,\phi}=\tilde C_{1,0,1}>0$. The Christoffel symbols,
\(C_{ijk}\) (or \(\Gamma^{i}{}_{jk}\)), are the numerical factors that
describe these necessary corrections. They quantify how much each
basis vector's components must change when transported along another
basis vector's direction while maintaining length and direction.

This leads to a physical insight. Let us (again) think of the vector
$\vec v$ to be parallel-transported as some momentary velocity. If we
take the curve parameter \(s\) to represent time on the clock of an
observer moving on that `world line' trajectory through space-time,
then the quantity \(dv_i/ds\) corresponds to the rate of change of a
velocity-vector's coordinates over time -- it is a coordinate of an
acceleration vector. In an observer-focused coordinate system,
i.e. one in which the spatial coordinates describe (some form of)
displacement from the observer's position, the important
transport-vector is ``forward in time'', with coordinates $(1,0,0,0)$
in that observer-focused $(t,x,y,z)$ coordinate-system.

This is how General Relativity describes gravitational
acceleration. An astronaut in orbit is in free-fall, following a
geodesic through spacetime. From their perspective, they feel no
forces and no acceleration. They are ``weightless'' because they are
on the straightest possible path.

However, we, observing from our Earth-fixed coordinate system, see
their path as a curve. The Christoffel symbols, which are non-zero in
our coordinate system, manifest as a coordinate-acceleration. We
interpret this acceleration as the force of gravity.

The ``fictitious forces'' we experience in a rotating frame (like the
Coriolis force) and the force of gravity are, in this geometric view,
the same kind of phenomenon. They are describing straight-line motion
(a geodesic) through a curved or accelerated coordinate system. This
idea is explored further in appendix~\ref{app:ChristoffelPhysics}.

Our next task is to find a way to compute these Christoffel
symbols. It seems like we have introduced a complicated new object,
but helpfully, the Christoffel symbols are not independent of other
information about the geometry of our space. At any given point, they
can be calculated directly from the metric tensor \(M\) and its
linearized rate-of-change w.r.t. position-coordinates. This means that
once we know how to measure distances (the metric), we automatically
know how to define straight lines (via the Christoffel symbols).

\section{Computing Parallel Transport Coordinate Corrections}\label{sec:computingChristoffel}

So what is the link between the metric and the Christoffel symbols? The
connection is based on the single physical principle called metric
compatibility. It is the simple idea that when we parallel transport two
vectors, the scalar product between them must not change.

This makes sense. Parallel transport is the mathematical definition of
``not turning'' a vector. If we take two vectors, \(\vec{v}\) and
\(\vec{w}\), and slide them both along a path without turning either of
them, then their lengths should not change, and the angle between them
should not change. Since all the information about lengths and angles
is available from the scalar product,

\begin{equation}
\vec{v} \cdot \vec{w} = \sum_{i,j} M_{ij} v_i w_j,
\end{equation}
it must remain constant during parallel transport.

Here is the challenge: if our coordinate system is non-Cartesian (like
polar coordinates) or if the geometry of spacetime is curved, our
basis vectors themselves can change from point to point. We see this
very clearly when we can refer to an objective notion of parallelism,
such as when describing the flat plane in polar coordinates:
The radial basis vector \(\vec{e}_r\) always
points radially outward, and the angular basis
vector \(\vec{e}_\theta\) is always tangent to the circle around the
origin through the present point. As we move from one point to
another, the directions of these basis vectors change.

This presents a problem. If our basis vectors are changing, how do we
know a vector's components are not changing just because of what the
basis vectors do, but the vector itself would actually be
parallel-transported? The affine connection provides the answer. It
gives us the mathematical ``correction factors'' that tell exactly how
the basis vectors change as we take a small step in our
space(-time). By including them in our equations, we can obtain a
quantitative description of ``parallel transport'' -- alongside a
prescription for how to relate local vector-coordinates given in one
coordinate system to another -- such that across different choices of
coordinates, there is agreement on the geometric meaning of ``parallel
transport of a vector over a small distance''.

This requirement that the scalar product remains constant despite the
changing basis vectors, together with a ``zero torsion in the vacuum''
requirement that is satisfied in General Relativity, is enough to give
us a recipe for calculating the Christoffel symbols from the
metric. While the full derivation is a bit tedious
(see appendix \ref{app:Christoffel}), the resulting formula is:

\begin{equation}C_{ik\ell} = \frac{1}{2} \sum_m (M^{-1})_{im} \left( \frac{\partial M_{mk}}{\partial x_\ell} + \frac{\partial M_{m\ell}}{\partial x_k} - \frac{\partial M_{k\ell}}{\partial x_m} \right).\end{equation}

This expression looks strange since we deduced a rather simple
property of a curved-space coordinate system -- the ``parallel
transport coordinate correction'' for a vector -- from a conceptually
higher-level quantity: the rate-of-change of the local
scalar-product-in-terms-of-vector-coordinates function. Here, the
(coordinate-)scalar product only cares about pairs of vectors, while
(coordinate-)parallel-transport cares about single vectors.

For completeness, we note that in the literature, this equation is
often stated in different form, with symbols ($\Gamma\leftrightarrow
C, M\leftrightarrow g$), and using upper/lower index placement on
$\Gamma$ and $g$ to indicate contravariant/covariant transformation
under local coordinate system change -- an important detail which
however only plays a minor role in our present story.

\begin{equation}\Gamma^i{}_{k\ell} = \frac{1}{2} \sum_m (g^{-1})^{im} \left( \frac{\partial g_{mk}}{\partial x^\ell} + \frac{\partial g_{m\ell}}{\partial x^k} - \frac{\partial g_{k\ell}}{\partial x^m} \right)\end{equation}

Each entry \(C_{ik\ell}\) is calculated using two ingredients:

\begin{itemize}
\tightlist
\item
  The terms \(\frac{\partial M_{mk}}{\partial x_{\ell}}\),
  \(\frac{\partial M_{m\ell}}{\partial x_{k}}\), and
  \(\frac{\partial M_{k\ell}}{\partial x_{m}}\), which describe how the
  coordinates of the metric tensor \(M\) vary as we move around in space,
  using our coordinate system.
\item
  The inverse of metric \((M^{-1})_{im}\), which acts like an inverse
  matrix in a linear equation. It converts the information about how the
  coordinates of a scalar product with a parallel-transported vector
  (a `vector-socket object', so to speak)
  change under parallel transport to information about how vector-coordinates
  themselves (a `vector-plug object') change under parallel transport.
\end{itemize}

A full example calculation of the Christoffel symbols can be found in
appendix~\ref{app:ChristoffelExample}.

In a raytracing context, we need to re-evaluate these
parallel-transport corrections at each step along a trajectory. If we
are given the geometry in terms of a ``metric tensor coordinates as a
function of position-coordinates'', JAX allows us to obtain a function
that calculates these correction-coefficients via an algorithm
transformation. The input-algorithm itself here may well itself have
been obtained from ``tangent vector length-squared as a function of
position-coordinates'' via another such automatic algorithm transformation.

\section{Geodesics as Autoparallel Curves}\label{sec:geodesics}

Now we can finally combine all our concepts to achieve a major goal:
finding an equation for the ``straightest possible paths,'' the
geodesics. The logic is beautifully direct:

By definition, a geodesic is a path that parallel-transports its own
tangent vector. This means the path does not ``turn'' relative to its
own direction of travel.

Let us describe our path by a (smooth) coordinate-tuple valued function
$\mathbb{R}\to\mathbb{R}^4: s\mapsto x_i(s)$.
So, if \(s\) is the parameter along the path, the ``rates of change of the coordinates
w.r.t. the path-parameter'' \(c_i = \frac{d x_i}{ds}\) is the
(curve-parameter-)velocity. At any given point on the curve, this
is a vector at that point.

The rule for parallel transporting a vector \(\vec{v}\) along the
momentary tangent-vector to this path
\(\vec{c}\) is the equation we established:

\begin{equation}
\frac{d v_i}{ds} = \sum_{j,k} \left(-C_{ijk}\right) \, v_j \, c_k
\end{equation}

For a geodesic, the vector being transported is the path's own tangent
vector. So, we set \(\vec{v} = \vec{c}\). This gives us:

\begin{equation}
\frac{d c_i}{ds} = -\sum_{j,k} C_{ijk} \, c_j \, c_k
\end{equation}

Now, let us substitute back in what \(\vec{c}\) is. The velocity is
the first derivative of position with respect to the curve parameter
(\(c_i = \frac{d x_i}{ds}\)), which means its rate of change
is the second derivative of position-coordinates
(\(\frac{d c_i}{ds} = \frac{d^2 x_i}{ds^2}\)). This gives us:

\begin{equation}
\frac{d^2 x_i}{ds^2} = -\sum_{j,k} C_{ijk} \, \frac{d x_j}{ds} \, \frac{d x_k}{ds}
\end{equation}

This is the ``geodesic equation''. Structurally, this is a coupled
system of second-order Ordinary Differential Equations (ODEs). For
our 2d map, we get two coupled equations (one for each value
of the index \(i\)). For the 3+1-d spacetime of General Relativity
(three spatial and one time dimension), we get four.

The geodesic equation may look complex, but its meaning is simple: It
shows how position (\(x_i\)) and (curve-parameter-)velocity (\(\frac{d
x_i}{ds}\)) define the (curve-parameter-)coordinate-acceleration for
an object in free fall (\(\frac{d^2 x_i}{ds^2}\)). Knowing the
acceleration at every moment allows us to calculate the entire
path. For our purpose -- raytracing -- it is appropriate to use some
numerical ODE solver in order to obtain a trajectory from a given
initial position, speed, and direction.

\section{Numerical Differential Equations}\label{sec:ode}

We want to numerically solve the ODE that describes geodesic motion by
iteratively advancing in small curve-parameter increments.
The simplest conceivable approach to numerically integrating such an ODE
would correspond to processing one short curve-parameter interval
\([s_n \ldots s_{n+1}]\) after another, and using
\(\Delta s_n:=s_{n+1}-s_n\), compute the motion-state at the next step
as:

\begin{strip}
\begin{equation}\begin{array}{lll}
&(t_{n+1},x_{n+1},y_{n+1},z_{n+1};\\
&\phantom((dt/ds)_{n+1},(dx/ds)_{n+1},(dy/ds)_{n+1},(dz/ds)_{n+1})\phantom)&\\
=&(t_n+\Delta s_n\cdot (dt/ds)_n,\ldots,z_n+\Delta s_n\cdot (dz/ds)_n;\\
 &\phantom((dt/ds)_n+\Delta s_n\cdot(d^2t/ds^2)_n,\ldots,
  (dz/ds)_n+\Delta s_n\cdot(d^2z/ds^2)_n)\phantom)
\end{array}
\end{equation}
\end{strip}

Here, the second derivatives are computed from the current motion-state
using the geodesic equation. This amounts to working out the local
rate-of-change of every motion-state parameter and incrementing that
parameter by
\(\{\text{curve parameter interval length}\}\cdot\{\text{rate of change}\}\).
In the limit of curve parameter interval lengths going to zero, we would
expect to get convergence to the actual dynamics.

In practice, such a basic scheme leads to major problems due to poor
numerical performance w.r.t. `quality of predictions vs. computational
effort'. The problem is that in general, the rate-of-change function
depends on its input motion-state vector, and approximating motion
over a time interval by multiplying the velocity at the start of that
interval with the length of the interval may be a poor estimate to the
quantity of actual interest -- the \emph{average} rate-of-change over
the given parameter interval when following the true trajectory, times
the length of that interval.

If we used such a blunt approach to ODE-integrate circular motion over time,
we would, for every time step \(\Delta t\), advance position \(\vec r\)
by \(\vec v\Delta t\), where \(\vec v\) is the momentary velocity --
which is tangent to the circle. So, at every time step, we are briefly
moving in a straight line ``as if the cord keeping us move in a circle
were loosened between the start and the end of the time interval''.
Motion would noticeably spiral outward, but halving the time-step would
reduce the deviation incurred at every step to roughly \(1/4\). Still,
since we then need twice as many steps, we only reduce the deviation by
50\% in magnitude by doubling computational effort.

More useful approaches are based on the following idea: at a given
point in motion-state space, we evaluate the rate-of-change function
not only once, but multiple times at a set of strategically-chosen
nearby positions, and use that information to determine a good-quality
approximation to how the rate-of-change function itself changes over
the integration parameter interval. If that approximation is such that
it would be correct for an order-$k$ polynomial dependency of the
rate-of-change on its inputs, we typically can expect that halving the
integration step size reduces the deviation for this interval to
$(1/2)^{k+1}$ (in alignment with the order of the most important
neglected correction), so the total deviation for a trajectory of now
twice as many integration steps by $(1/2)^k$. With such an improved
approach, we can then obtain a far better approximation to the true
solution of the ODE with given computational effort.

Here, we will use the Runge-Kutta RK4 integration scheme as a
``cookbook'' recipe for getting a good estimate of the average
rate-of-change over a time interval. Using JAX, our main tool for
implementing a numerical ODE solver is the \texttt{jax.lax.scan()}
function that we here employ to collect the to-be-recorded outputs of
a \((\{\text{current state}\},\{\text{curve parameter interval length}\})\mapsto(\{\text{next state}\},\{\text{state to be recorded}\})\)
function in an array. In our use case, \(\{\text{next state}\} = \{\text{state to be recorded}\}\).

\begin{figure*}[ht]
\centering
\includegraphics[width=0.8\textwidth]{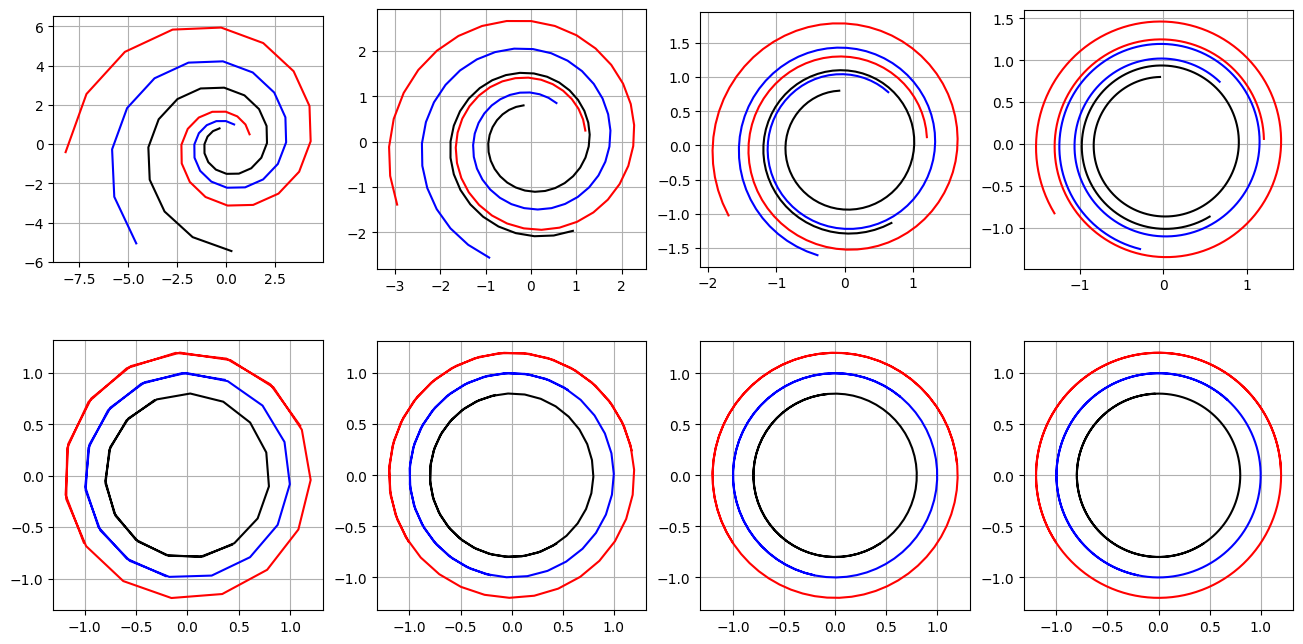}\\
\begin{lstlisting}[language=Python]
def rk4_estimate(f, y0, ds):
  k1 = f(y0)
  k2 = f(y0 + (0.5 * ds) * k1)
  k3 = f(y0 + (0.5 * ds) * k2)
  k4 = f(y0 + ds * k3)
  return (k1 + 2.0 * k2 + 2.0 * k3 + k4) / 6.0

def simple_estimate(f, y0, ds): return f(y0)

def ode_solve(estimate_func, f, y0, s_samples):
  def inner_func(y_now, s_step):
    rate = estimate_func(f, y_now, s_step)
    y_next = y_now + s_step * rate
    return y_next, y_next
  y_final, ys_collected = jax.lax.scan(inner_func, y0, jnp.diff(s_samples))
  return ys_collected

jax_ode_solve = jax.jit(ode_solve,
                        static_argnames=('f', 'estimate_func'))

circular = jax_ode_solve(
  rk4_estimate,  # Alternative: simple_estimate
  (lambda x: jnp.concatenate([-x[1:2], x[:1]], axis=0)),
  jnp.array([1.0, 0.0]), jnp.linspace(0.0, 10.0, 201))
\end{lstlisting}
\caption{Comparing ODE-integration of $\omega=1$ circular motion
  via naive approach (top row) and via RK4 (bottom row) when subdividing
  the time interval `\texttt{[0..10]}` into 25, 50, 100, 200 steps.}
\label{fig:ode-comparison}
\end{figure*}

Our solver is implemented in such a way that we can provide an
arbitrary ``estimate the average rate of change over a curve-parameter
interval'' function that implements the integration scheme. We show
code and illustrate the performance of the RK4 approach in comparison
to the naive approach in Figure~\ref{fig:ode-comparison}.

\section{Aspects of Raytracing in a Black Hole Geometry}\label{sec:raytracing}

Having developed the tools of parallel transport, geodesics, and
numerical integration, we now turn to the specific setting in which
our raytracer will operate. We first generalize our treatment from
curved two-dimensional surfaces to the four-dimensional curved
space-time of General Relativity, and then show how to extract the
metric tensor from a line element, the key input our raytracer needs.

While this might look like a big step, all the machinery we developed
so far carries over in a straightforward way, if we pay attention to
one important detail: so far, we could introduce `time' as a
curve-parameter in order to talk about local velocities. Upon close
inspection of the construction, we find that naming the curve
parameter `time' was merely a storytelling device to develop the
mathematical machinery, using the mathematical notion of kinetic
energy as a key bridge towards introducing a scalar product on curved
spaces. In space-time, time is one of our coordinates, so we have to
be more careful in discriminating between curve-parameters (and
rate-of-change of some quantity with respect to a curve-parameter
change) and `time'. In any case, we can here always hold on to
explicit constructions, climbing down to code and even numbers in
order to clarify how any specific statement made is to be interpreted.

\subsection{From Curved Surfaces to Curved Space-Time}

In our ``curved space-time ray-tracing'' setting, a light ray is the
trajectory of a fictitious ``photon'' particle that moves along a
geodesic. Its momentary motion-state is described by its space-time
position coordinates \((t,x,y,z)\) as well as its momentary coordinate
rate-of-change \(((dt/ds),(dx/ds),(dy/ds),(dz/ds))\) that sets the
direction.

The ``local scalar product'' in general relativity is such that, when
picking some point ${\mathcal P}$ in space-time, we can find the
coordinates of the metric tensor at that point, $M^{({{\mathcal P}})}_{ij}$,
and then introduce some local coordinate system for
velocity vectors at ${\mathcal P}$ such that the basis vectors of that
coordinate system are mutually-perpendicular to one another. We find
that we even can do this in such a way that three of these basis
vectors have length-squared 1 with respect to that local scalar
product. If we also could do this for the fourth basis vector, we
would describe some four-dimensional space in which we can freely
rotate any coordinate-direction into any other such direction. Since
this is not how space-time geometry works, the scalar product of the
fourth mutually-orthogonal basis vector with itself cannot be
positive. Since there is a physical notion of ``larger or smaller
distance along the fourth direction, time'', it also is not
zero. Hence, it must be negative, and we can choose the length of the
time-directed basis vector to be such that it has scalar product $-1$
with itself.

Generically speaking, geometry studies the interplay between
transformations of some space (such as rotations of 3d space) and the
objects that are invariant under these transformations (like the angle
between two vectors). Contemplating such a `generalized scalar
product' of two vectors that is symmetric and linear in both
arguments, but can take on both positive and negative values, may seem
unusual at first, in particular since it will allow us to find vectors
with non-zero coordinates whose generalized length-squared
nevertheless is zero. Concretely, if $S(\cdot, \cdot)$ is such a
scalar product and we have two vectors $\vec u, \vec v$ satisfying
$S(\vec u, \vec v)=0$, $S(\vec u, \vec u)=+1$, $S(\vec v, \vec v)=-1$
(as would be the case for the $x$-directed and $t$(ime)-directed basis
vectors $\vec e_x$ and $\vec e_t$), we would find that for
$\vec w:=\vec u+\vec v$, we have
$S(\vec w, \vec w)=S(\vec u,\vec w)+S(\vec v,\vec w)$ (due to linearity).
Hence, $S(\vec w, \vec w)=S(\vec u, \vec u)+0+S(\vec v,\vec v)=1-1=0$.

If we now consider applying coordinate-transformations to this basis
of velocity-vectors, this does not affect the geometrical meaning of
this (generalized) scalar product: there is consensus among users of
different coordinate systems on whether some particular vector is
`length-squared zero' or not. Moreover, if we only consider
coordinate-transformations that keep basis vectors mutually-orthogonal
and normalized to length-squared $\pm 1$, this excludes -- for example
-- `stretching only one direction of 3d space', but retains 3d
rotations, plus additionally some extra transformations that `mix'
space- and time-coefficients of velocity vectors. These can be
understood to correspond to transitioning from the orthonormal
coordinate basis of one observer to an orthonormal coordinate basis of
another observer who is not at rest relative to the first -- for what
the first observer considers `moving a small step forward along its
time coordinate but not changing its space-coordinate' will in the
second observer's description involve a small step that adjusts both
its time-coordinate and spatial coordinates. These extra so-called
``boost'' transformations enlarge the 3d spatial rotations to
``geometrically sensible rotations of 3+1-dimensional space-time''.
If some velocity-vector has length-squared zero, there is agreement on
this fact across all observers' coordinate systems, and as such, those
velocity-vectors are suitable to model tangents to the trajectories of
photons, i.e. light rays. Parallel transport then tells us how to
extend a trajectory from a given starting point and initial
velocity-vector -- which, as explained, is constrained to have
length-squared zero for a photon. In this approach, the
curve-parameter of the `light ray' trajectory should be thought of as
an auxiliary mathematical tool without much of a direct physical
interpretation.

For flat, not-curved space-time (i.e. there is no matter or energy
that could be the source of space-time curvature), we can pick
`Minkowski coordinates' $(t,x,y,z)$ such that the coordinate-entries
of the metric tensor everywhere are $M_{00}=-1,
M_{11}=M_{22}=M_{33}=+1$, with all other $M_{ij}=0$. In the presence
of space-time curvature, this will no longer work so nicely, and we
would in particular expect that if we have some solid ball that causes
gravity, then outside that ball, we would expect the
position-dependent metric tensor to be such that everywhere on a
concentric sphere enclosing the ball, the Christoffel symbols would
describe a coordinate-correction for ``moving forward in time,
starting on that sphere with zero velocity in radial direction'' that
decreases the radial coordinate over time, i.e. ``space-time curvature
is such that free-falling bodies accelerate towards the ball''.

Albert Einstein -- and almost at the same time, David Hilbert --
worked out how the precise relationship between matter and geometry
has to look like back in
1915~\cite{einstein1915feldgleichungen,hilbert1915grundlagen}. For the
scenario described above, Karl Schwarzschild solved these equations in
1916~\cite{schwarzschild1916gravitationsfeld} and presented the line
element for any spherically symmetric distribution of mass in its
exterior region. We nowadays understand this equation to also describe
the limiting case of a (non-rotating) black hole.

In our language, the ``Schwarzschild Line Element'' can be presented
in the form shown below. Here, $\theta, \phi$ are `latitude' and
`longitude' coordinates on any sphere around the center of gravity,
the ``radial'' coordinate $r$ is defined such that any large shell of
radius $r$ has surface area $4\pi\,r^2$, as in flat space, and the
``time'' coordinate $t$ is such that for an observer very very far
from the source of gravity, who moves in such a way that
$(r,\theta,\phi)$ do not change for them over time, time passes as in
free space, and for some fixed-$(r_P,\theta_P,\phi_P)$-place $\tilde
P$ (three coordinates, not four!) closer to the origin, `being at time
$t_*$ at $\tilde P$' means the following: If a light signal got sent
from an at-rest (defined as above) place $\tilde {\mathcal Q}$
having coordinates
$(r_{\mathcal Q},\theta_{\mathcal Q}=\theta_{\mathcal P},\phi_{\mathcal Q}=\phi_{\mathcal P })$ with $r_{\mathcal Q}\gg r_/{\mathcal P}$, which
started at $\tilde {\mathcal Q}$-time $t^{({\mathcal Q})}_-$ such that it reaches $\tilde {\mathcal P}$
just at $t_*$, then gets mirrored back up towards $\tilde {\mathcal Q}$ where it
then arrives a $\tilde {\mathcal Q}$-time $t^{({\mathcal Q})}_+$, then the time-coordinate
for the observer staying at place $\tilde {\mathcal Q}$ for the (temporal)
midpoint between sending and receiving the signal also has
time-coordinate $t_*$. Here, $(t,r,\theta,\phi)$ are
position-coordinates\footnote{Here, the angle $\theta$ is measured
from the North Pole, not the equator (like latitude would be).},
and their rates-of-change with curve-parameter $s$ for some suitably
chosen collection of curves is $t'=d(t(s))/ds,\,x'=d(x(s))/ds$, etc.

\begin{equation}
  \begin{array}{lcl}
    |\vec v|^2_\text{Schwarzschild}&=&-F(r)\cdot (t')^2 + (r')^2/F(r)\\
    &&+ r^2\cdot(\theta')^2+r^2\sin^2(\theta)\cdot(\phi')^2
  \end{array}
\end{equation}

We get Schwarzschild's solution of Einstein's differential equations
of space-time curvature if we pick $F(r)=1-\frac{r_S}{r}$, where $r_S$
is the so-called `Schwarzschild radius' -- the higher the total amount
of mass $M_\text{total}$, the larger this parameter. Intuitively,
`something goes wrong in this description if we set $r=r_S$'. While
every radially-outgoing light ray that starts at $r>r_S$ can reach
arbitrarily high $r$ as time advances, ``the maths goes wrong'' if we
try to start at $r<r_S$. So, $r_S$ is the threshold radius above which
outgoing light signals ``can reach infinite distance from the
gravitating body''. Notably, in the equation above, the summands not
involving $F(r)$ align perfectly with ``measuring distance when only
moving along the surface of a sphere of the given radius''.

If we pick this as our `background space-time geometry' and want to
build a raytracer, it is useful to not actually work in spherical
coordinates but instead consider ``auxiliary Cartesian coordinates''
that only work for $r>r_S$ and are defined such that they satisfy the
natural flat-space relations between ``spherical'' and ``Cartesian''
coordinates, $(x,y,z)=(r\sin\theta\cos\phi, r\sin\theta\sin\phi,
r\cos\theta)$. The physical meaning of the ``Schwarzschild line
element'' then is that if we want to compute the ``generalized
length-squared'' of a vector $(t',x',y',z')$ and at a point with
coordinates $(t, x, y, z)$, we have to work out $r=\sqrt{x^2+y^2+z^2}$
and be careful about splitting the vector into `radial', `spatially
normal-to-radial' and `temporal' components, where the
normal-to-radial contribution to the length-squared is as in flat
space, the `temporal' contribution receives a $r$-dependent
correction-factor that makes one unit step of coordinate-time
correspond to less physical time passing the closer we are to the
center, and the `radial' contribution receives a correction factor
that goes into the opposite direction.

The spatial correction-factor makes a shell $[r-\Delta r/2; r+\Delta
r/2]$ have a 3d volume of not $V\approx 4\pi r^2\Delta r$ for
$0<\Delta r/r\ll 1$, but a \emph{larger} spatial volume. In other
words, if we consider an asteroid of radius 100 meters, then when
taking gravity into account, an upward-pointing ladder for climbing
between a sphere around its center of surface area $4\pi\cdot (1\,{\rm
km})^2$ and a sphere of surface area $4\pi\cdot (1.010\,{\rm km})^2$
will turn out to be \emph{a tiny bit longer} than $10\,{\rm
m}$. Still, since the limit-from-above $\lim_{r_-\to
r_S}\,\int_{r_-}^{r_+}\,dr/\sqrt{1-r_S/r}$ converges, lowering a rope
all the way down to the Schwarzschild radius would require a larger,
but still finite, amount of rope in comparison to the situation in
flat-space geometry.

Figure~\ref{fig:jax-schwarzschild} shows a JAX implementation of the
Schwarzschild solution in such a form that is especially well suited
for raytracing.

\begin{figure*}[ht]
\begin{lstlisting}[language=Python]
def schwarzschild_L2_func(txyz, d_txyz, r_schwarzschild=1.0):
  xyz = txyz[1:]
  r = jnp.linalg.norm(xyz)
  e_r = xyz / r
  part_e_r = (d_txyz[1:] @ e_r) * e_r
  part_perpendicular = d_txyz[1:] - part_e_r
  ds2_perpendicular = jnp.square(part_perpendicular).sum()
  s_factor = jnp.where(r > r_schwarzschild, 1 - r_schwarzschild / r, jnp.nan)
  ds2_radial = jnp.square(part_e_r).sum() / s_factor
  ds2_time_abs = jnp.square(d_txyz[0]) * s_factor
  return ds2_radial + ds2_perpendicular - ds2_time_abs
\end{lstlisting}
\caption{The Schwarzschild ``line element'' in Cartesian-from-polar
  coordinates, as a JAX-function.}
\label{fig:jax-schwarzschild}
\end{figure*}

\subsection{Putting it All Together}

Via numerical ODE-integration, we can obtain numerical sample-points
along geodesics. In earlier sections, we developed the computational
machinery for obtaining parallel-transport corrections from a
``line-element function'' given not by some algebra, but here
implemented as computer code. We have seen how the
autoparallel-transport ODE looks on paper.
Figure~\ref{fig:geodesic-ode} shows how to transform a JAX-function
that numerically computes the Christoffel symbols into a JAX-function
that gives us the rate-of-change function to plug into the ODE-integrator.

Apart from this, we have to handle a few mundane details with rather
straightforward code which we do not show here. On these, we refer to
the appendix with the fully self-contained basic raytracer, where the
implementation paid attention to properly documenting the construction
in alignment with industry best practice (albeit not making use of
Python static type annotations, and accepting that mutable state such
as JAX compilation caches are owned by objects in global scope). These
bits and pieces need to cover:

\begin{itemize}
  \item At a given ``eye-point'' observer position away from the
    gravitational center, introducing a proper ``physically
    orthonormal'' local space-time basis.
  \item Converting vector-coefficients that refer to that basis to
    vector-coefficients that refer to a set of four
    ``coordinate-line'' curves through the observer-point along each
    of which only one of the four space-time-coordinates changes (and
    also is the curve-parameter), while the others remain constant.
  \item For each point on a screen that sits just in front of the
    observer towards the direction of the gravitational center,
    finding the initial tangent-vector of the geodesic that connects
    the eye-point to that point on the screen. This initial
    tangent-vector must correspond to ``moving at lightspeed'' (so,
    have scalar product zero with itself w.r.t. the local metric at
    the observer), but go ``backwards in time'' -- since we are
    answering the question where the photon came from that hit the
    camera at $t=0$ and came from that direction.
  \item vector-parallelizing ODE-integration to apply always the same
    calculations to different numerical data, doing the computations
    for multiple pixels all at the same time in a way that can exploit
    GPU parallelism.
  \item Determining the direction of the light ray at a
    sufficiently-late ODE integration endpoint and mapping that to
    some fictitious ``map of the sky'' on the celestial sphere.
\end{itemize}

With JAX, parallelization is rather straightforward -- at least as
long as we are not delving into some subtler details about array-data
device placement optimizations.

JAX uses a higher level approach than earlier machine learning
accelerators (such as for example TensorFlow~\cite{tensorflow2016}) that are
primarily designed to allow convenient specification of generic
tensor-algebraic constructions as they often arise in machine learning
applications. Rather, the primary design objective of JAX is to make
it straightforward to specify algebraic transformations of
tensor-algebraic constructions. One obvious such transformation that
powers practically all of contemporary machine learning is
``reverse-mode automatic differentiation~\cite{speelpenning1980compiling}'',
which maps an algorithmic description of a numerical calculation to an
efficient algorithmic description of a numerical calculation of the
gradient of the output w.r.t. input-parameters. A conceptually simpler
such transformation is `vectorization' of code, i.e. mapping a
computation to another computation where each input (and output)
carries an additional ``number of the example'' array-index that
performs the same calculations on all examples in parallel.
This is what the \texttt{jax{.}vmap()} function allows us to do.
Being able -- as we did here -- to focus on the construction
for a single trajectory and then add hardware-supported parallelization
as an afterthought clearly is a very appealing feature of JAX.

W.r.t. `deciding how to color the pixel, given the ODE-integration
trajectory', the most basic approach is to consider a very large
screen very far behind the gravitational center with a checkerboard
pattern on it, and determine from the direction of the outgoing light
ray whether it will hit a light or dark square. A perhaps somewhat
more interesting alternative (or complementary) idea would be to
analyze each recorded photon trajectory to find equator-crossings, and
(via interpolation) determine the ``first'' (w.r.t. integration-step)
such crossing for which the radius is inside some given interval. This
allows us to render the gravity-distorted image of some ``saturn
rings'' type feature around the gravitational center. If we do not
eliminate trajectories that come closer to the center than some `size
of the celestial object' radius but allow going down all the way to
the Schwarzschild radius $r_S$, we obtain a rendered image of a black
hole. Some black holes are known to have ``rings'' around them -- more
precisely, an ``accretion disc'', but these discs generally rotate at
relevant speeds around a black hole that itself carries angular
momentum\footnote{We are here not discussing in depth what it actually
means for empty space to ``rotate''.}. In such a setting, the
``Schwarzschild line element'' geometry is not adequate to describe
the situation, and a more complex geometry such as the ``Kerr-Newman
metric''~\cite{kerr1963gravitational,newman1965note}
should be used instead.

The code release that accompanies this work\footnote{At \url{https://github.com/paradigms-of-intelligence/raytracing-black-holes-with-jax}}
goes well beyond the objective of this article to build a bridge between ML technology and
physics by describing a skeletal framework in detail and shows how to
render an image of a rotating black hole with a physically reasonable
model for an accretion disc (this involves modeling temperature and
then also thermal radiation as a function of distance from the black
hole) in Kerr geometry, showing phenomena such as the increase of the
effective visual area of the disc due to gravitational light-bending,
`Einstein ring' type distortion of the image of stars behind the black
hole, as well as effects due to gravitational redshift and
relativistic beaming.

\begin{figure*}[ht]
\begin{lstlisting}[language=Python]
def geodesic_dy_dt_func_from_christoffel_func(christoffel_func):
  def dy_dt_func(motion_state_y):
    c_pos, c_tangent = jnp.unstack(motion_state_y.reshape(2, -1), axis=0)
    rate_tangent = -jnp.einsum('ijk,j,k->i', christoffel_func(c_pos),
                               c_tangent, c_tangent)
    return jnp.concatenate([c_tangent, rate_tangent], axis=0)
  return dy_dt_func


def geodesic_dy_dt_func_from_metric_func(metric_func):
  christoffel_func = christoffel_func_from_metric_func(metric_func)
  return geodesic_dy_dt_func_from_christoffel_func(christoffel_func)    
\end{lstlisting}
\caption{Obtaining the ``rate of change of motion-state'' function for ODE-integration from the line element}
\label{fig:geodesic-ode}
\end{figure*}

\section{On Curvature}\label{sec:curvature}

So far, we only took Schwarzschild's solution of the Einstein
equations from the literature, pondering neither whether it actually
is a ``no gravitational sources in the exterior region'' solution as
claimed, nor how Einstein's gravitational field equations work.

Explaining the physical interpretation of Einstein's equations,
i.e. ``how matter curves space-time'' is beyond the scope of this
work. On that topic, readers might find e.g. John Baez's
article~\cite{baez2005meaning} useful. Here, we nevertheless want to
explore one useful detail: using JAX-based numerics to verify that a
solution is indeed a vacuum solution, i.e. source-free. We
nevertheless need to sketch some properties of the theory to at least
allow for a high level understanding of how the elements of the
construction fit together.

In our approach, we focus attention on space-time geometry, and for
any given geometry, Einstein's equations can be used to calculate what
the mass-density at every point in space-time would have to be for
matter to create some space-time geometry like this. As a special
relativity perspective immediately suggests here, mass-density, while
relatable to everyday intuition, must be part of a more complex
entity. When changing the description of some physical situation from
what one observer sees to what another observer sees who is in
relative motion to the first, `energy' and `momentum' parameters of a
moving object mix, and also, `forward in time' world-line tangents
also acquire spatial components from the perspective of the other
observer -- due to the relativity of simultaneity.

Mass-density is ``the amount of energy crossing through a spatial 3d
spatial window in 3+1d space-time around the point in question, per
window-volume''. This then must be understood as ``energy-or-momentum
flux through some small 3d window (not necessarily containing only
spatial directions) around the space-time point in focus''. This `4d
energy-momentum flux' is called the `stress-energy tensor', and if we
decide (which conceptually might be a mistake) to describe the 3d
window's size and orientation as a 4d vector\footnote{The argument
here is a ``factorization'' argument: Any function or property of this
``window'' that matters for determining spacetime curvature source
strength can be expressed symbolically to depend on the specific shape
of the window, but any such function $\{\text{window}\}\mapsto
f(\{\text{window}\}$ can be re-written in terms of $f
= \{\text{window}\}\mapsto f^*(W(\{\text{window}\})$, where $W(\cdot)$
extracts this 4d-vector and $f^*$ is a variant of $f$ that operates on
that data.}
that is perpendicular to the window whose
length is proportional to its volume, we can describe this entity in
terms of a (space-time position dependent) $4\times 4$ matrix of
coefficients that contain information not only about what local
energy-density any observer passing through this point would find, but
also information about the energy-flux or
momentum-density\footnote{These two quantities have to be the same --
in electrodynamics, this is the energy-flux `Poynting vector'
$\propto \vec E\times\vec B$, for example.}, as well as the mechanical
stress tensor.

This ``stress-energy tensor'' with coordinates $T_{ij}$ satisfies a
conservation law: energy and momentum can flow from one point to
another, but the total outflux of energy or momentum out of any
4-dimensional space-time volume always must be zero. So, if these
``sources of space-time curvature'' somehow can be equated to a
geometric quantity that is related to curvature, this must be a
quantity which for geometric reasons satisfies an extra constraint
that aligns with this ``volume integral equals zero'' property. If we
furthermore demand that ``twice as much curvature equals twice as much
energy-momentum flux'', i.e. want to have some ``linear
theory\footnote{Just as springs ultimately will exhibit nonlinear
behavior under strong tension, it is natural to wonder whether there
may be quadratic-in-curvature or even higher order corrections to
General Relativity. Experimentally, the existence of such quadratic
corrections would be very difficult to establish via measurements, and
on the theory side, there are constraints from quantum gravity
(whatever it might look like) not allowing any predicted probabilities
to be negative. Superstring theory as one candidate for a possible
embedding of general relativity into a quantum theory would predict
such corrections.} of space-time elasticity'', this fixes the
structure of the theory almost completely. The only remaining
questions then are about ``how do we fix a conversion factor so that
we recover Newton's Gravitational Law in the small-curvature limit''
and the value of the cosmological constant.

One way to think about local curvature is in terms of local holonomy:
for a given space-time point, we look at all small 2d patches of
space-time that contain that point, stay in its neighborhood, and have
a smooth perimeter. Transporting an orthonormal coordinate system
around that small perimeter-loop will induce a small rotation of the
coordinate system where, for small loops, only the size and
orientation of the loop matter, but not its shape\footnote{Also here,
we see a factorization argument at work: the small-rotation-from-loop
function uniquely factorizes through an
effective-size-and-orientation-of-small-loop function.}.  It turns out
that both ``size and orientation of a small loop'' and ``small amount
of 4d rotation'' can here be described by 6-coordinate geometric
quantities, and `curvature' then is a linear relation between these,
so can be described by a local $6\times 6$-matrix. Deeper explorations
reveal that we can always make it such that this matrix is symmetric,
implying that local curvature can be described by not $6\times 6=36$
but in fact no more than $6\cdot 7/2=21$ parameters. It furthermore
turns out that there is one extra property that can be
interpreted\footnote{ One might at first think that such a
`vector=vector' identity would introduce four more constraints. Closer
investigation reveals that when taking the other constraints into
account, we only get a not-already-satisfied condition from picking
four independent basis vectors, so in dimension $D$, we get
$\binom{D}{4}$ constraints -- in space-time dimension $3+1=4$ we see
one nontrivial constraint.}  as admitting a straightforward way to
obtain the correction to a vector $\vec a$ when parallel-transporting
around a small loop formed by vectors $\vec b, \vec c$ from the
corrections to vector $\vec b$ when parallel-transporting around a
small loop formed by vectors $\vec a, \vec c$ and to $\vec c$ when
parallel-transporting around a loop formed by $\vec a,\vec b$.  So,
local curvature in space-time actually only has 20 independent
parameters. Since local energy-momentum flux can be described by no
more than $4\times 4=16$ parameters\footnote{Actually, we only have 10
independent parameters, since one can show that this matrix will be
symmetric}, it is immediately clear that the relation will have to be
such that some linear function of the curvature-parameters will equal
some linear function of the energy-momentum flux.

It is also clear that ``local curvature'' will somehow be computed
from parallel transport (of a coordinate frame around a small loop),
and one can intuitively understand that for small loops, this will
have to involve the Christoffel symbols as well as their first
derivatives. In the usual presentation of the theory, local curvature
is not described by a symmetric $6\times 6$ matrix: Even if we could
reshape the information in that form, it is normally presented as a
$(4,4,4,4)-$array, the ``Riemann Tensor'', with many extra properties
(or constraints) that reduce the number of independent parameters from
$4^4=256$ to the aforementioned $20$.

It turns out that, when starting from some arbitrary proposed
space-time geometry, by far not every setting where local geometry
always has one time-direction is associated with a physically
meaningful scenario: It may well turn out that the computation tells
us that for the given geometry, local energy density would have to be
negative\footnote{Whether quantum gravity could allow some such thing
is a complex discussion.} in some regions, for example.

Here, we want to focus on computational aspects and content ourselves
with taking the formulae for computing the Riemann tensor from the
Christoffel symbols as well as for computing the momentum-energy flux
(called ``stress-energy tensor'') from the Riemann tensor from the
literature. Aligning with our conventions so far, this reads:

\begin{equation}
  \begin{array}{lcl}
    R_{ijk\ell}&=&\frac{\partial C_{ij\ell}}{\partial x_k}-\frac{\partial C_{ijk}}{\partial x_\ell}+\\
    &&+\sum_{m}\left(C_{ikm}C_{mj\ell}-C_{i\ell m}C_{mjk}\right)\\
    &&\\
    G_{ij}&=&\sum_m R_{mimj}-\\
    &&-\frac{1}{2}M_{ij}\sum_{m,k,\ell}\left(M^{(-1)}\right)_{k\ell}R_{mkm\ell}\\
    &&\\
    T_{ij}&=&G_{ij}/(8\pi G_\text{Newton})
  \end{array}
\end{equation}

So, for a solution to be a vacuum solution, $G_{ij}=0$ needs to hold
for every $(i,j)$ index-pair at every point where we evaluate local
curvature\footnote{The situation would naturally be more involved for
charged black holes: due to their charge, they have to have an
electromagnetic field around them, which ``stores energy'', so that
field's energy shows up as a source of gravity.}.

As e.g. M. Visser's introduction to the Kerr metric
explains~\cite{visser2007kerr}, even for the
geometry of a rotating black hole, establishing $G_{ij}=0$ in the
exterior region is a major algebraic challenge that only few
professional physicists mastered to do with pen and paper. Resorting
to symbolic algebra shows that even for a computer, the calculation is
not so simple that one would not get measurable benefit from switching
to clever parametrizations. This problem is a useful setting for
explaining the key difference between doing theoretical physics via an
entirely symbolic approach vs. employing the complementary ``physical
constructions are represented as differentiable computer algorithms''
approach we present in this work. In the latter
differentiable-algorithms approach, we lose equational reasoning, so
no longer can easily establish equivalence of two parameter-dependent
expressions for all values of the parameters, and furthermore are
forced to work with limited numerical precision (often, but not
necessarily, dictated by hardware). We still can collect strong
evidence for the equivalence of two expressions given as algorithms,
such as by evaluating them at a large number of probe points and then
comparing values. The biggest benefit of the ``differentiable
algorithms'' approach may well be that we are guaranteed that the
price of a local gradient is ``bounded and cheap'', i.e. if we have
some nonlinear $\mathbb{R}^m\to\mathbb{R}^n$ function that can be
computed in such a way that we can remember all intermediate
quantities in the calculation, we can always compute the Jacobian with
extra effort no larger than by a factor $k\cdot \text{min}(m, n)$,
where $k=8$ is certainly attainable, and in practice one often finds
$k\sim 4-5$. The symbolic approach more readily leads to an explosion
in the number of terms due to generally less diligent bookkeeping of
re-use of intermediate quantities. The actual algorithm-transformation
naturally also requires computational effort, but this is linear in
the size of the algorithm. Applying this machinery to problems of
finding geodesics on low-dimensional spaces is possible (as this work
demonstrates), but the true benefits of ``algorithm differentiation''
show for problems that have far higher-dimensional gradients -- for
some advanced ML applications, these can have even many billions of
entries.

A ``cartoon-level'' description of
the problem might be that the symbolic approach generally computes
$(f(x)\cdot g(x)\cdot h(x))'=f'(x)\cdot g(x)\cdot h(x)+f(x)\cdot g'(x)\cdot h(x)+f(x)\cdot g(x)\cdot h'(x)$
and then -- this is the important part -- typically tries to perform
simplifications that create entirely new terms which do not share much
sub-structure with already existing terms).
In contrast, algorithmic differentiation
would internally represent the calculation in the perhaps unusual-looking form
$(a\mapsto(b\mapsto(c\mapsto ((d\mapsto a\cdot d)(b\cdot c)))))(f(x))(g(x))(h(x))$,
and then when taking the derivative transforms this\footnote{The
transformation shown here actually is forward-mode automatic
differentiation. Reverse-mode automatic differentiation is far more
efficient for $\mathbb{R}^m\to\mathbb{R}^n$ algorithm-functions with
$m\ll n$ but not so easy to express in this language.}
to the formidable-looking $((a,a')\mapsto((b,b')\mapsto((c,c')\mapsto
(((d,d')\mapsto a'\cdot d+a\cdot d')(b'\cdot c+b\cdot
c')))))(f(x),f'(x))(g(x),g'(x))(h(x),h'(x))$, which however has the
key advantage that each of $f(x)$, $g(x)$, $h(x)$ only occurs once in
the expression, even post-substitution of these functions by some
complex definitions.

Figure~\ref{fig:einstein} shows code for obtaining the Riemann tensor\footnote{
What we here call $R_{ijk\ell}$ is actually $R^i{}_{jk\ell}$ in the relativity literature,
where index placement indicates that $i$ is an ``output'' (``plug'')
vector-coordinate index, while the other indices are ``input'' (``socket'')
vector coordinates: When using the Riemann tensor the way we introduced it,
$j,k,\ell$ bind against coordinate-indices of the vector to be taken
around a small loop and the vectors defining the small patch, while $i$
is an ``output'' coordinate-correction. A $R_{ijk\ell}$ in the relativity literature
would correspond to $\sum_m M_{im} R_{mjk\ell}$ in our approach.
This also explains why in the formula to compute $G_{ij}$, there has
to be a $M^{(-1)}$: while the sum over $m$ connects a ``plug-index'' to
a ``socket-index'', we need a ``geometric double-plug cable'' type object to connect
two ``socket-indices''.}
$R_{ijkl}$ and Einstein tensor $G_{ij}$ from the Christoffel symbols
and ultimately the line element.  Figure~\ref{fig:schwarzschildG}
shows how to numerically evaluate the Einstein tensor for one generic
point. As expected for a vacuum solution, this is zero -- to numerical
accuracy.

\begin{figure*}[ht]
\begin{lstlisting}[language=Python]
def riemann_func_from_christoffel_func(christoffel_func):
  jacobian_christoffel_func = jax.jacobian(christoffel_func)
  def riemann_func(coords_pos):
    christoffel = christoffel_func(coords_pos)
    d_christoffel = jacobian_christoffel_func(coords_pos)
    # We use jnp.einsum() throughout, also for simple transposition.
    aux = (jnp.einsum('ijlk->ijkl', d_christoffel) +
           jnp.einsum('ikm,mjl->ijkl', christoffel, christoffel))
    # The whole expression is aux_ijkl - aux_ijlk.
    return aux - jnp.einsum('ijkl->ijlk', aux)
  return riemann_func


def einstein_func_from_L2_func(L2_func):
  metric_func = metric_func_from_L2_func(L2_func)
  christoffel_func = christoffel_func_from_metric_func(metric_func)
  riemann_func = riemann_func_from_christoffel_func(christoffel_func)
  def einstein_func(coords):
    metric = metric_func(coords)
    riemann = riemann_func(coords)
    ricci = jnp.einsum('mimj->ij', riemann)
    return (ricci - 0.5 * jnp.einsum('ij,ij->', ricci, jnp.linalg.inv(metric))
            * metric)
  return einstein_func
\end{lstlisting}
\caption{Obtaining the ``Riemann'' and ``Einstein'' Tensor-functions from the line element.}
\label{fig:einstein}
\end{figure*}

\begin{figure*}[ht]
\begin{lstlisting}[language=Python]
schwarzschild_einstein_func = einstein_func_from_L2_func(schwarzschild_L2_func)
jax_schwarzschild_G = jax.jit(schwarzschild_einstein_func)
generic_coords = jnp.array([5.5, 2.3, -3.0, 0.7])
einstein1 = jax_schwarzschild_G(generic_coords)
print(einstein1.round(12))
# Prints:
# [[-0.  0.  0.  0.]
#  [ 0. -0.  0.  0.]
#  [ 0. -0.  0. -0.]
#  [ 0. -0. -0.  0.]]
\end{lstlisting}
\caption{Calculating the Einstein Tensor at some generic point in Schwarzschild geometry}
\label{fig:schwarzschildG}
\end{figure*}

\section{Images}\label{sec:images}

In its most basic form, the raytracing code produces an image of a
black hole in front of a checkerboard background. In order to get a
better idea of the geometry -- and also to align with astronomical
features that some black holes are known to have -- we also add a
small extension to the basic renderer that adds an equatorial ring
around the black hole. Conceptually, this is somewhat of a sideline to
our main story, despite producing interesting physics, and as such, we
keep the code adjustment separate from the core construction.

In our approach, JAX is used to render image-patches, and more basic
\texttt{numpy}-based post-processing is used to then translate
geodesic data into image information. For such data-processing, one
would normally want to not exercise the (inefficient) CPython bytecode
interpreter, but instead rely on vectorized \texttt{numpy} ufuncs. In
order to add the checkerboard, we need to work out whether each ray
successfully cleared the hole and determine its direction at a faraway
point where space-time is approximately flat again. In order to add
the ``accretion disc'' annulus, we need to find, for each ray, the
first-along-the-trajectory equator-crossing (keeping track of
direction in order to color the upper and lower side differently) for
which the coordinate-radius is inside the selected interval for the
annulus.

\begin{figure*}[ht]
  \centering
  \[
  \begin{array}{cc}
    \includegraphics[width=0.4\textwidth]{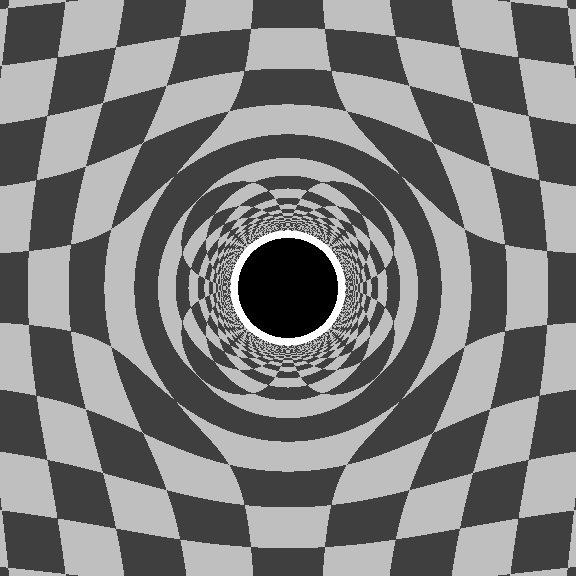}&
    \includegraphics[width=0.4\textwidth]{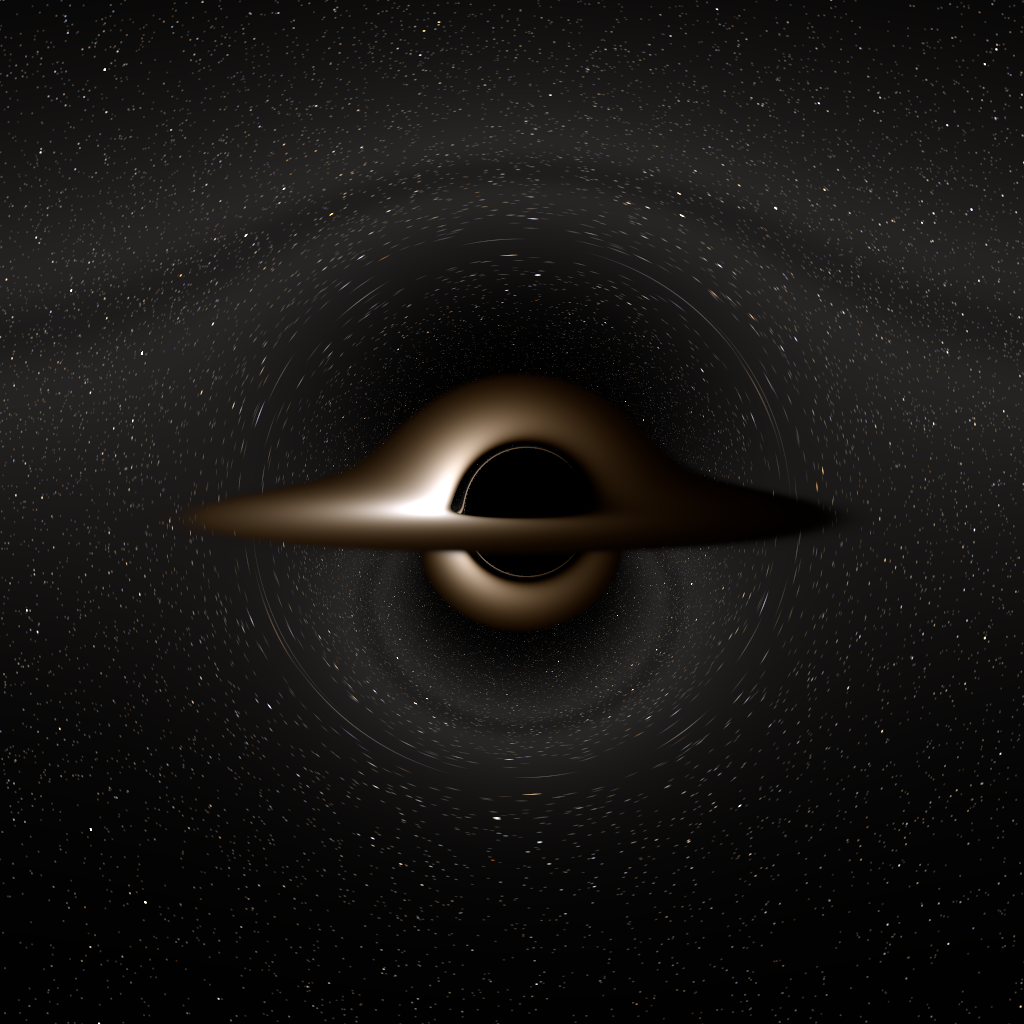}\\
    \includegraphics[width=0.4\textwidth]{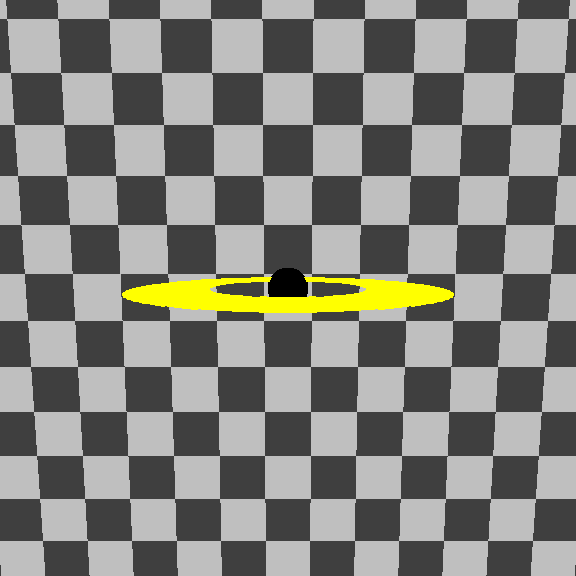}&
    \includegraphics[width=0.4\textwidth]{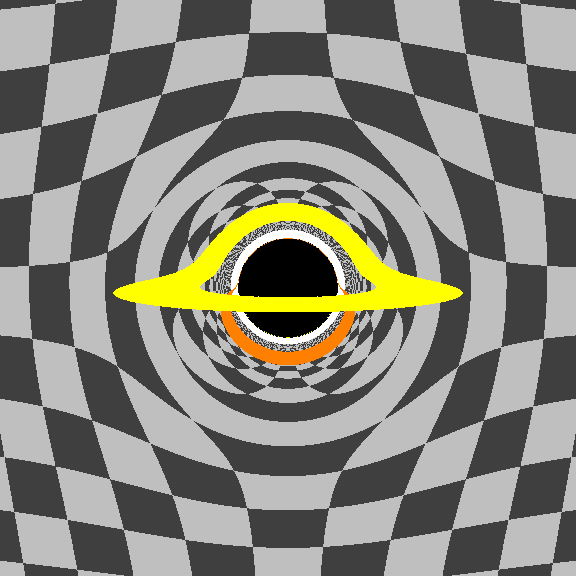}
  \end{array}
  \]
  \caption{Top left: Light deflection around a black hole in front of a checkerboard background.
    Bottom left: Zero-mass black object with disc around it; surface area equals black
    hole event horizon area. Bottom right: the black hole with a disc around it.
    Top right: A spinning black hole with accretion disc in front of a star field.}
\label{fig:BlackHoleImages}
\end{figure*}

Figure~\ref{fig:BlackHoleImages} shows three images produced by the
raytracer as presented here, plus a fourth one generated by a more
evolved implementation available from the repository accompanying this
work. The top left image shows a black hole in front of a faraway
checkerboard pattern behind it and was generated with the raytracer as
presented here. As of 2026, producing such a $576\times 576$ image
takes under six minutes using a free-of-charge Google Colab notebook
virtual machine~\cite{bisong2019google} with T4 GPU
acceleration\footnote{Computational efficiency could be improved a lot
by e.g. replacing Runge-Kutta RK4 ODE-integration with a more
sophisticated method, or even just using different step sizes close to
and away from the black hole. This has not been done here in the
interest of keeping the code compact and readily
understandable.}. Black pixels indicate light rays that in the
discretized stepping hit the interior $r<r_{\text{Schwarzschild}}$
forbidden region. White pixels indicate light rays that failed to go
past the hole -- intuitively, they ``show the white screen behind the
photographer''. In this set-up, the center of the black hole is
visually in front of the center of a black checkerboard square, whose
center gets distorted to a ring -- an ``Einstein Ring'' -- via
``gravitational lensing'': Light rays that reach the camera under a
vision angle corresponding to this ring started out parallel to the
camera-axis well behind the black hole. The four corners of this
central checkerboard square each show up twice on the perimeter of the
roughly-annular region -- once on the inside, once on the
outside. Light rays passing closer to the black hole than this get
deflected at high angles, reaching 90 degrees right at the boundary of
the white annulus. A higher resolution image would show that further
images of the checkerboard appears even closer to the black hole, due
to light rays that circle around the hole once, twice, etc.

The bottom left image shows the same background, but with the camera
moved a bit upward, and with the black hole replaced with a zero-mass
black sphere whose surface area equals the black hole's horizon area
$4\pi\cdot r_S^2$ -- implemented by replacing $1-r_S/r$ with $1-0\cdot
r_S/r$ in the code of \texttt{schwarzschild\_L2\_func()}. (One notices
that, due to the change in perspective, vertical lines on the
checkerboard now converge towards the bottom of the image.) This
sphere is surrounded by an equatorial disc that reaches out from the
surface of a sphere with area $4\pi\cdot\left(4\cdot r_S\right)^2$
to the surface of a sphere with area $4\pi\cdot\left(8\cdot
r_S\right)^2$. The bottom right image shows the exact same setting,
but now with the black sphere replaced with a black hole. Light rays
reaching the upper side of the disc are shown with yellow pixels,
light rays reaching the lower side of the disc are shown with orange
pixels. In order to eliminate some minor numerical artefacts that tend
to arise mostly due to large ODE integration step size, the number of
elements of \texttt{ode\_steps} has been increased from 401 to
1201. Since in the bottom-left image, the sphere's center is in front
of a \emph{light} square, the ``Einstein ring tile'' here also is
light, not dark as in the top left image. One notices two orange
streaks going through the white-pixels region. These are due to light
rays that pass closer to the hole than the inner edge of the disc, go
around the hole and hit the accretion disc from below while going
towards the observer.

Finally, the top right image shows the capabilities of an extended
implementation (available on github) of the raytracer. It replaces the
Schwarzschild geometry with the Kerr-Newman metric in Kerr-Schild
Cartesian coordinates, which describes a black hole with angular momentum.
($a/M=0.6$, $Q=0$). The fixed-step RK4 integrator is replaced by a
Dormand-Prince 5(4) scheme with adaptive step-size control, which is
more robust near the strong-field region. On top of these structural
changes, the extended code adds several visual and physical features:
a thin accretion disk with Novikov-Thorne radiative flux, a background
star field loaded from an ESO Milky Way panorama, physically-based
blackbody color rendering via Planck's law and CIE color matching
functions. The asymmetric brightness of the disc is due to Doppler
shift and relativistic beaming.

\section{The Physics of General Relativity}\label{sec:grphysics}

Our journey so far has given us an approach to use computing in order
to understand and, should we want to do so, even quantitatively check
some predictions of General Relativity such as the amount of starlight
deflection around the sun during a solar eclipse~\cite{dyson1920}.

Still, this so far is only part of the story. In particular, we took
the ``black hole geometry'', i.e. ``the Schwarzschild solution'' as a
``prediction from General Relativity'' literature result without
pondering in depth three key questions:

\begin{itemize}
\item First, why we are \emph{forced} to model gravity as space-time curvature
  if we want to retain the ``equivalence principle'',
\item Second, what the local physical relations are that govern space-time curvature and
  ``the sources of gravity'', such as especially mass(/energy)-density,
\item Third, how the Schwarzschild solution arises as a solution to these ``field equations''.
\end{itemize}

In the interest of brevity, we can here only sketch the underlying
ideas and insights that provide answers to these questions.

Special Relativity is, at its core, a geometric theory of space-time
in which ``vacuum lightspeed'' is understood not so much as a physical
property of photons, but a geometric property of space-time itself. It
then is a prediction that particles with rest mass zero must move at
this speed, but that in itself does not make a claim about
photons\footnote{We could construct a model world in which the photon
would have a tiny mass. It then would not move at ``causality
threshold speed'', and if we were to live in such a world, we likely
would use some such term rather than ``vacuum lightspeed''.}

A key experimental observation about gravity is that it affects all
masses the same way, irrespective of their composition. In principle,
we could define the mass of an object in terms of its inertia --
i.e. as ``the amount of force that needs to act on it for one second
in order to increase its velocity by one meter per second''.
Alternatively, we could try to define mass in terms of ``how strong
the attractive gravitational force is that pulls the object towards
earth''. It is at first not clear at all that these two definitions
are compatible! If, for example, we had two metal balls, one made of
magnesium and one made of gold, which according to this approach have
identical inertial mass, why would we expect their gravitational mass
to also align? Would it not be conceivable that, say, neutrons were to
feel gravity a bit stronger than protons, and the higher proportion of
neutrons in the gold nucleus then would make the gold ball feel
gravity a bit more strongly in relation to its inertial mass?
Experiments that try to find a difference between ``inertial'' and
``gravitational'' mass can be designed to be extremely sensitive and
all failed to find a difference. This makes it plausible that there
might be a deeper reason why there fundamentally cannot be a
difference. This would be a natural consequence of ``gravity'' and
``inertia'' being the same phenomenon, i.e. motion under the influence
of gravity being the same as inertial motion. For that idea to work,
gravitational acceleration would have to be a ``fictitious force''
just like the force an astronaut suddenly experiences when their
spaceship turns on its rocket engines and undergoes acceleration. A
loose object in the cockpit would retain its zero-gravity free-fall
motion, but from the perspective of the now-upward-accelerating
cockpit coordinate frame, it would ``fall down''. But if this were the
correct explanation, we obviously would struggle to explain why, from
the perspective of a Londoner, ``things in Sydney fall in a direction
they call \emph{up}''. Pondering this situation more deeply shows that
the ``equivalence principle'' (equivalence between gravitation and
inertia) plus the principle of locally-constant speed of light that is
the foundation of special relativity, when taken together, pretty much
force us to accept that gravity must be related to the non-flatness of
classical space-time.

\section{Conclusion and Outlook}\label{sec:conclusion}

As we hope to have shown with these explanations, ML accelerator
libraries such as JAX can open many doors when we think about them
more broadly -- not just as tools used in the construction of complex
ML / AI systems, but as convenient and powerful enabling tools also
for turning rather nontrivial physics based constructions into fast
code that can utilize GPU acceleration rather easily. While more
advanced uses may run into problems with the underlying Python
programming language's somewhat misaligned focus as a
mostly-easy-to-use scripting language rather than a precision tool for
specifying mathematical constructions with stringency, the popularity
of Python certainly is a benefit here.

If a construction like this works and is shown to have major
advantages for explaining General Relativity -- clearly not the
simplest of physical theories -- this makes it very plausible that
similar approaches that emphasize making the physics tangible by
representing physical quantities not as differentiable symbolic
expressions, but differentiable algorithms, could be used to great
effect when trying to make other theories more tangible as well.

Here, one should think not only about educational purposes but also
about actually studying situations that are messy to investigate with
symbolic methods alone. Two important fields that immediately come to
mind here are wave optics and antenna theory: With JAX and some
guidance, it should be well within reach for an undergraduate physics
student to build a basic finite difference time domain Maxwell
equation solver all on their own.

It must be pointed out that the underlying ideas and theory have been
around for much longer than the Deep Learning revolution -- or even
the earlier GPU revolution that made access to impressive numerical
computational power very affordable. In that light, availability of ML
accelerator frameworks that turn out to also have such interesting
other uses is a fortunate accident that greatly accelerated progress
on algorithmic differentiation, elevating it from a somewhat obscure
but economically very valuable niche one can build an academic career
in to a widely used and understood commonplace piece of engineering
mathematics.

This article is part of a wider effort to build bridges between
different disciplines via showing how to enlarge their shared language
for expressing ideas. In the view of the authors, this is especially
important for two reasons: First, due to generative AI having made
access to (some non-human form of) advanced expertise much more
affordable, the relative value of overly narrow specialization saw
some decline, and the ability to talk across discipline boundaries is
becoming more important than before. Second (and from direct own
experience of one author), different advanced fields of study that
have to manage a high level of complexity of their constructions often
evolved different approaches to handling that complexity -- such as to
identify problems with a construction early -- that in principle could
be carried over and used to great benefit in other disciplines as
well. A key impediment is a lack of shared language that would
facilitate carrying over such ideas. To give a concrete example,
computer scientists, type theorists, ab initio physics researchers and
quantum field theorists all have evolved ways to talk about
identifying impossible-since-misaligned precise relations between
entities. Yet, we are still far from the point where a ``physical
dimension'' factor on a quantity would be widely (and across
disciplines) understood as being equivalent to a dedicated range-1
array/tensor-index indicating transformation under an irreducible
representation of an abelian scaling symmetry. Allowing advanced
disciplines to reap the benefits of deep insights evolved by other
such disciplines will require far more bridge-building work than what
this article can contribute.

Circling back specifically to General Relativity research, even if we
do not consider generalizing use of JAX beyond raytracing, the methods
presented here should be applicable rather directly to -- for example
-- problems concerning efficient image generation for higher
dimensional objects, such as those that have been described
in~\cite{hertog2019imaging}.

Looking at other efforts to make General Relativity accessible, many
texts make a deliberate choice to de-emphasize the role of
coordinates, focusing the discussion on abstract mathematical concepts
such as tangent and cotangent spaces, sections, pullbacks,
Maurer-Cartan forms, etc. In contrast, our supremely coordinate-heavy
approach obviously takes the exact opposite direction -- we here even
mostly avoid to talk about the difference between tangent and
cotangent vector coordinates. The abstraction-focused approach
provides much value, first and foremost in weeding out ad-hoc
constructions that would look plausible when knowing only about
coordinate-based numerics but are physically meaningless. In the
authors' view, reaching a deep level of understanding of the subject
requires mastering the entire ladder that connects abstract maths at
the one end to low level numerics at the other end. For different
students, the answer to the question which entry point to the topic
works best for them can be very different, and usually depends on
what they have prior expertise with.

\bibliographystyle{unsrt}
\bibliography{references}

@incollection{bisong2019google,
  title={Google colaboratory},
  author={Bisong, Ekaba},
  booktitle={{Building} machine learning and deep learning models on google cloud platform: a comprehensive guide for beginners},
  pages={59--64},
  year={2019},
  publisher={Springer}
}

@misc{jax2018,
  author = {James Bradbury and Roy Frostig and Peter Hawkins and Matthew James Johnson and Yash Katariya and Chris Leary and Dougal Maclaurin and George Necula and Adam Paszke and Jake Vander{P}las and Skye Wanderman-{M}ilne and Qiao Zhang},
  title = {{JAX}: composable transformations of {P}ython+{N}um{P}y programs},
  url = {http://github.com/jax-ml/jax},
  version = {0.3.13},
  year = {2018},
}

@incollection{pytorch2019,
  title = {{PyTorch}: {An} {Imperative} {Style}, {High-Performance} {Deep} {Learning} {Library}},
  author = {Paszke, Adam and Gross, Sam and Massa, Francisco and Lerer, Adam and Bradbury, James and Chanan, Gregory and Killeen, Trevor and Lin, Zeming and Gimelshein, Natalia and Antiga, Luca and Desmaison, Alban and Kopf, Andreas and Yang, Edward and DeVito, Zachary and Raison, Martin and Tejani, Alykhan and Chilamkurthy, Sasank and Steiner, Benoit and Fang, Lu and Bai, Junjie and Chintala, Soumith},
  booktitle = {Advances in Neural Information Processing Systems 32},
  pages = {8024--8035},
  year = {2019},
  publisher = {Curran Associates, Inc.},
  url = {http://papers.neurips.cc/paper/9015-pytorch-an-imperative-style-high-performance-deep-learning-library.pdf}
}

@inproceedings{tensorflow2016,
  author = {Mart{\'\i}n Abadi and Paul Barham and Jianmin Chen and Zhifeng Chen and Andy Davis and Jeffrey Dean and Matthieu Devin and Sanjay Ghemawat and Geoffrey Irving and Michael Isard and Manjunath Kudlur and Josh Levenberg and Rajat Monga and Sherry Moore and Derek G. Murray and Benoit Steiner and Paul Tucker and Vijay Vasudevan and Pete Warden and Martin Wicke and Yuan Yu and Xiaoqiang Zheng},
  title = {{TensorFlow}: {A} {System} for {Large-Scale} {Machine} {Learning}},
  booktitle = {12th USENIX Symposium on Operating Systems Design and Implementation (OSDI 16)},
  year = {2016},
  isbn = {978-1-931971-33-1},
  address = {Savannah, GA},
  pages = {265--283},
  url = {https://www.usenix.org/conference/osdi16/technical-sessions/presentation/abadi},
  publisher = {USENIX Association},
  month = nov
}

@article{baez2005meaning,
  title={The meaning of {Einstein}’s equation},
  author={Baez, John C and Bunn, Emory F},
  journal={American journal of physics},
  volume={73},
  number={7},
  pages={644--652},
  year={2005},
  publisher={American Association of Physics Teachers}
}

@article{einstein1915feldgleichungen,
  title={{D}ie {F}eldgleichungen der {G}ravitation},
  author={Einstein, Albert},
  journal={Sitzungsberichte der K{\"o}niglich Preu{\ss}ischen Akademie der Wissenschaften},
  pages={844--847},
  year={1915}
}

@article{hilbert1915grundlagen,
  title={Die {Grundlagen} der {Physik}.({Erste} {Mitteilung}.)},
  author={Hilbert, David},  
  journal={Nachrichten von der Gesellschaft der Wissenschaften zu G{\"o}ttingen, Mathematisch-Physikalische Klasse},
  volume={1915},
  pages={395--408},
  year={1915}
}

@article{dyson1920,
  author = {Dyson, F. W. and Eddington, A. S. and Davidson, C.},
  title = {{A} {Determination} of the {Deflection} of {Light} by the {Sun}'s {Gravitational} {Field}, from {Observations} {Made} at the {Total} {Eclipse} of {May} 29, 1919},
  journal = {Philosophical Transactions of the Royal Society of London Series A},
  volume = {220},
  pages = {291--333},
  year = {1920},
  doi = {10.1098/rsta.1920.0009}
}

@article{bruneton2020,
  author = {Bruneton, Eric},
  title = {{Real-time} {High-Quality} {Rendering} of {Non-Rotating} {Black} {Holes}},
  journal = {arXiv preprint arXiv:2010.08735},
  year = {2020},
  eprint = {2010.08735},
  archivePrefix = {arXiv},
  primaryClass = {cs.GR}
}

@article{vincent2011,
  author = {Vincent, F. H. and Paumard, T. and Gourgoulhon, E. and Perrin, G.},
  title = {{GYOTO}: a new general relativistic ray-tracing code},
  journal = {Classical and Quantum Gravity},
  volume = {28},
  pages = {225011},
  year = {2011},
  doi = {10.1088/0264-9381/28/22/225011},
  eprint = {1109.4769},
  archivePrefix = {arXiv}
}

@article{bronzwaer2018,
  author = {Bronzwaer, T. and Davelaar, J. and Younsi, Z. and Mo{\'s}cibrodzka, M. and Falcke, H. and Kramer, M. and Rezzolla, L.},
  title = {{RAPTOR}. {I}. {T}ime-dependent radiative transfer in arbitrary spacetimes},
  journal = {Astronomy and Astrophysics},
  volume = {613},
  pages = {A2},
  year = {2018},
  doi = {10.1051/0004-6361/201732149},
  eprint = {1801.10452},
  archivePrefix = {arXiv}
}

@book{isaacson2007einstein,
  title     = {Einstein: {His} {Life} and {Universe}},
  author    = {Isaacson, Walter},
  year      = {2007},
  publisher = {Simon \& Schuster},
  address   = {New York, NY},
  isbn      = {978-0-7432-6473-0}
}

@article{kerr1963gravitational,
  title={Gravitational field of a spinning mass as an example of algebraically special metrics},
  author={Kerr, Roy P},
  journal={Physical review letters},
  volume={11},
  number={5},
  pages={237},
  year={1963},
  publisher={APS}
}

@article{newman1965note,
  title={Note on the {Kerr} spinning-particle metric},
  author={Newman, Ezra T and Janis, Allen I},
  journal={Journal of Mathematical Physics},
  volume={6},
  number={6},
  pages={915--917},
  year={1965},
  publisher={American Institute of Physics}
}

@article{schwarzschild1916gravitationsfeld,
  title={{\"U}ber das {Gravitationsfeld} eines {Massenpunktes} nach der {Einsteinschen} {Theorie}},
  author={Schwarzschild, Karl},
  journal={Sitzungsberichte der k{\"o}niglich preussischen Akademie der Wissenschaften},
  pages={189--196},
  year={1916}
}

@book{schweber2020qed,
  title={QED and the men who made it: {Dyson}, {Feynman}, {Schwinger}, and {Tomonaga}},
  author={Schweber, Silvan S},
  year={2020},
  publisher={Princeton University Press}
}

@book{sussman2015structure,
  title={Structure and {Interpretation} of {Classical} {Mechanics}},
  author={Sussman, Gerald Jay and Wisdom, Jack},
  year={2015},
  publisher={Mit Press}
}

@book{speelpenning1980compiling,
  title={Compiling fast partial derivatives of functions given by algorithms},
  author={Speelpenning, Bert},
  year={1980},
  publisher={University of Illinois at Urbana-Champaign}
}

@article{visser2007kerr,
  title={The {Kerr} spacetime: A brief introduction},
  author={Visser, Matt},
  journal={arXiv preprint arXiv:0706.0622},
  year={2007}
}

@article{hertog2019imaging,
  title={Imaging higher-dimensional black objects},
  author={Hertog, Thomas and Lemmens, Tom and Vercnocke, Bert},
  journal={Physical Review D},
  volume={100},
  number={4},
  pages={046011},
  year={2019},
  publisher={APS}
}

@article{mulvey1980nature,
  title={The {Nature} of {Matter}, {Wolfson} {College} {Lectures}, 1980},
  author={Mulvey, John H},
  journal={The Nature of Matter, Wolfson College Lectures},
  year={1980}
}

\appendix

\onecolumn

\section{Plotting Flight Paths}\label{app:plotflightpaths}

The code in this appendix shows how to plot a path that consists of
straight line segments in the (latitude, longitude) coordinate system
in terms of what they would look like under stereographic
projection\footnote{We deliberately use the \texttt{basemap} Python
package and not more modern alternatives, since this package by now
only rarely gets updates that might break user code.}.

\begin{lstlisting}[language=Python]
# For the analogy involving map projections, the 'basemap' library is used
# to provide a familiar visual reference for non-Cartesian coordinates.
# !pip install basemap

import itertools, jax, numpy, matplotlib.pyplot, scipy.interpolate
from mpl_toolkits import basemap   # Requires `pip install basemap`
from jax import numpy as jnp
jax.config.update("jax_enable_x64", True)


def get_map(axes):
  "Returns Basemap initialized for South Polar stereographic projection."
  bmap = basemap.Basemap(projection='spstere',
                         boundinglat=50,
                         lon_0=0,
                         resolution='l', ax=axes)
  bmap.drawcoastlines()
  bmap.fillcontinents(color='coral', lake_color='aqua')
  bmap.drawmapboundary(fill_color='aqua')
  bmap.drawparallels(numpy.linspace(-80., 80, 17))
  bmap.drawmeridians(numpy.arange(-180, 180, 36))
  # Adding a Cartesian grid overlay to show the map's (x,y) coordinates.
  x_min, x_max = axes.get_xlim()
  y_min, y_max = axes.get_ylim()
  for x in numpy.linspace(x_min, x_max, 11):
      axes.axvline(x, color='gray', linestyle='--', linewidth=0.75, zorder=1)
  for y in numpy.linspace(y_min, y_max, 11):
      axes.axhline(y, color='gray', linestyle='--', linewidth=0.75, zorder=1)
  return bmap

# Abbreviations for coordinate transformations back-and-forth
# between latitude/longitude and x/y map-plot coordinates - where
# our map size will be 10 cm x 10 cm, and coordinates are in cm.
MAP_SIZE_CM = 10


def xy_from_lat_lon(bmap, lat, lon):
  "Computes map-(x, y) from basemap, latitude, and longitude."
  x0, y0 = bmap(lon, lat)
  # Basemap is a bit unusual w.r.t. the role of '[xy]min' and '[xy]max'.
  return (MAP_SIZE_CM * (x0 - bmap.xmax) / (bmap.xmin - bmap.xmax),
          MAP_SIZE_CM * (y0 - bmap.ymax) / (bmap.ymin - bmap.ymax))


def xy_flight_path_from_ll_flight_path(bmap, lat_lon_from_t):
  def xy_from_t(t):
    # We are spelling out all the intermediate steps here.
    lat, lon = lat_lon_from_t(t)
    return xy_from_lat_lon(bmap, lat, lon)
  return xy_from_t


demo_path_lat_lon = numpy.array(
    [[-33.87, 151.21 - 360],   # Sydney
     [8.98, -79.52],           # Panama City
     [-34.60, -58.38],         # Buenos Aires
     [-33.92, 18.42],          # Cape Town
     [41.38, 2.17],            # Barcelona
     [4.17, 73.51],            # Male (Maldives)
     [35.68, 139.76],          # Tokyo
     [-33.87, 151.21]],        # Return to Sydney
    dtype=numpy.float64)
demo_path_ts = numpy.arange(demo_path_lat_lon.shape[0]) * 6 * 3600.0
demo_path_ll = scipy.interpolate.interp1d(demo_path_ts, demo_path_lat_lon,
                                          kind='linear', axis=0)
get_demo_path_xy = lambda bmap: (
    xy_flight_path_from_ll_flight_path(bmap, demo_path_ll))

fig, axes = matplotlib.pyplot.subplots(figsize=(24, 8), ncols=3)
ax_a, ax_b, ax_c = axes
ax_a.grid(); ax_b.grid()
ax_b.set_aspect('equal'); ax_c.set_aspect('equal')
ax_a.set_ylabel('Latitude'); ax_a.set_xlabel('Longitude')
bmap = get_map(ax_c)
flight_ts = numpy.linspace(0, demo_path_ts[-1], 1001)
demo_path_xy = get_demo_path_xy(bmap)
flight_lls = [demo_path_ll(t) for t in flight_ts]
ax_a.plot([lon for lat, lon in flight_lls],
          [lat for lat, lon in flight_lls], '-b')
ax_a.plot([lon for lat, lon in flight_lls[::50]],
          [lat for lat, lon in flight_lls[::50]], 'ob')
ax_b.set_xlabel('Map-X'); ax_b.set_ylabel('Map-Y')
flight_xys = [demo_path_xy(t) for t in flight_ts]
ax_b.plot([x for x, y in flight_xys],
          [y for x, y in flight_xys], '-k')
ax_b.plot([x for x, y in flight_xys[::50]],
          [y for x, y in flight_xys[::50]], 'ok')
x_map, y_map = bmap([lon for lat, lon in flight_lls],
                    [lat for lat, lon in flight_lls])
ax_c.plot(x_map, y_map, '-w', linewidth=3.5, label='Flight Path')
fig.savefig('/tmp/flight_path.pdf'); fig.show()
\end{lstlisting}

\twocolumn

\section{An Example of a Coordinate Transform and Scalar Product}\label{app:polar}

To get a better feel for how coordinate transforms affect the scalar
product, let us look at a simple example: converting from Cartesian
coordinates $(x, y)$ to polar coordinates $(r, \theta)$ in two
dimensions.

The transformation itself is pretty straightforward:
\begin{equation}
  x = r \cos(\theta), \qquad y = r \sin(\theta).
\end{equation}
In vector form, we can write the position vector $\vec{x}$ as a function
of the new coordinates:
$\vec{x}(r, \theta) = (r \cos(\theta),\; r \sin(\theta))$.

Next, we need the local basis vectors for our new coordinate system,
$\vec{e}_r$ and $\vec{e}_\theta$. We find these by taking the partial
derivatives of our position vector with respect to the new
coordinates. One notes that the coordinate-entries of these vectors
are given with respect to the original Cartesian coordinate system.

\begin{equation}
  \vec{e}_r
    = \frac{\partial \vec{x}}{\partial r}
    = (\cos\theta,\; \sin\theta),
  \qquad
  \vec{e}_\theta
    = \frac{\partial \vec{x}}{\partial \theta}
    = (-r\sin\theta,\; r\cos\theta).
\end{equation}
One can see that the basis vector $\vec{e}_\theta$ depends on the radial
coordinate~$r$. While $\vec{e}_r$ has a constant length of~1, the
length of $\vec{e}_\theta$ is not~1, it is~$r$.  This means that a
one-unit change in $\theta$ corresponds to a physical distance that gets
larger the farther one is from the origin. This is just like how one
degree of longitude covers a greater distance at the equator than it
does near the poles.

The metric tensor, $M$, is a matrix whose entries are the scalar
products of the basis vectors.  This lets us compute the ``geometrically
meaningful'' local scalar product in our new coordinate system.  The
formula for the entries is $M_{ij} = \vec{e}_i \cdot \vec{e}_j$.

Let us calculate each entry:
\begin{align*}
  M_{rr} &= \vec{e}_r \cdot \vec{e}_r
           = \cos^2\theta + \sin^2\theta = 1, \\
  M_{\theta\theta} &= \vec{e}_\theta \cdot \vec{e}_\theta
           = r^2(\sin^2\theta + \cos^2\theta) = r^2, \\
  M_{r\theta} &= M_{\theta r}
           = \vec{e}_r \cdot \vec{e}_\theta = 0.
\end{align*}
When we put it all together, the metric tensor looks like this:
\begin{equation}
  M = \begin{pmatrix} 1 & 0 \\ 0 & r^2 \end{pmatrix}.
\end{equation}
This matrix is diagonal because the basis vectors $\vec{e}_r$ and
$\vec{e}_\theta$ are locally orthogonal (perpendicular).  The second
diagonal entry is $r^2$ (and not~1) because the length of the
$\vec{e}_\theta$ vector depends on the radius~$r$.

Finally, let us use this matrix to find the scalar product of two
vectors, $\vec{v}$ and $\vec{w}$, in polar coordinates.  The formula is
$\vec{v}\cdot\vec{w} = \vec{v}^{\,T} M\, \vec{w}$.  If $\vec{v}$ has
components $(v_r, v_\theta)$ and $\vec{w}$ has components
$(w_r, w_\theta)$, we get:
\begin{equation}
  \vec{v} \cdot \vec{w}
    = v_r\, w_r + r^2\, v_\theta\, w_\theta.
\end{equation}
The scalar product is not just the sum of the products of the components
(e.g., $v_r w_r + v_\theta w_\theta$) as it would be in Cartesian
coordinates.  The $r^2$ factor is essential to account for the actual
geometry of the coordinate system.  This example shows why the metric
tensor is so important for correctly computing geometrically meaningful
quantities in non-Cartesian coordinate systems.

If, at a point with coordinates $(r, \theta)$, we are given a vector
with $(\vec e_r, \vec e_\theta)$ coordinates $(r', \theta')$, we can
use the scalar product to obtain the length-squared of that
vector. This gives us the line element for polar coordinates, from
which we can (as we have seen) go back to the metric via the
polarization identity:

\begin{equation}
  ds^2 = (r')^2 + r^2\cdot(\theta')^2.
\end{equation}

\section{Deriving the Connection Coefficients}\label{app:Christoffel}

When we work with a geometry where there is a local notion of lengths
and angles for tangent vectors, we usually (but not always) would
want parallel transport to be ``metric compatible''. This means that
if we take two arbitrary vectors
$\vec a^{({\mathcal P})}, \vec b^{({\mathcal P})}$ at
point\footnote{In General Relativity, this is a space-time point, also
  called an ``event''.} $\mathcal P$, and simultaneously
parallel-transport them along some given smooth curve, the rate-of-change
of the scalar product between these vectors always is zero -- and we
want this property to hold for every curve along which we parallel
transport.

Already by considering the situation in flat space, we notice that
metric compatibility alone is not strong enough to completely fix the
connection coefficients, i.e. the linear mapping from local
parallel-transport ``velocity vector'' to ``rate-of-change of vector
coordinate $j$ per vector coordinate $i$''. We could, for example,
imagine flat 3d space where parallel transport of a local tangent
vector by some distance $L$ along the $+z$-axis goes hand in hand
with counterclockwise rotation of that vector by an angle
$\alpha=\omega_0L$ for some fixed $\omega_0$. Since rotations preserve
lengths and angles, this parallel transport clearly would be ``metric
compatible''.
Since small rotations are of the form $\left(\begin{array}{cc}+\cos\varepsilon&-\sin\varepsilon\\+\sin\varepsilon&+\cos\varepsilon\end{array}\right)=I+\left(\begin{array}{cc}\phantom+0&-\varepsilon\\+\varepsilon&\phantom+0\end{array}\right)+{\mathcal O}(\varepsilon^2)$,
the ``momentary rate-of-change proportional to $\varepsilon$'' is an antisymmetric matrix.
In this example, where we have an underlying Cartesian coordinate system,
but with ``unusual'' parallel transport, this means that at every point, the $C_{ijk}$
are antisymmetric under exchange $i\leftrightarrow j$ -- since $\sum_k C_{ijk} d_k$ is
the angular velocity (expressed as an antisymmetric matrix) for rotation around the
z-axis as we move with momentary velocity-vector given by $\vec d$ -- where in this
example, only $d_2=``d_z''$ is relevant. This means that the only nonzero entries
of $C$ are $C_{012}=-C_{102}=\omega_0$.

Looking at our geodesic equation, which relates the rate-of-change of
the momentary velocity vector, i.e. the tangent vector to the geodesic
curve as we move along it, to the connection coefficients and the
velocity vector itself,
\begin{equation}
  (d/ds v_i(s)) = -\sum_{j,k}C_{ijk}\,v_j(s)\,v_k(s)
\end{equation}
it is clear that if we decompose $C$ into a part that is symmetric
under exchange $j\leftrightarrow k$ and one that is antisymmetric,
$C_{ijk}=C^S_{ijk}+C^A_{ijk}=\frac{1}{2}\left(C_{ijk}+C_{ikj}\right)+\frac{1}{2}\left(C_{ijk}-C_{ikj}\right)$,
only the symmetric part $C^S$ matters for the geodesic equation: the
antisymmetric part multiplies $v_jv_k$, which itself is symmetric under
this index exchange, so $C^A_{ijk}v_jv_k=-C^A_{ijk}v_jv_k=0$.

In our example, we have
$C^S_{ijk}=\frac{1}{2}\omega_0\bigl(
  \delta_{i0}\delta_{j1}\delta_{k2}+\delta_{i0}\delta_{k1}\delta_{j2}
  -\delta_{i1}\delta_{j0}\delta_{k2}-\delta_{i1}\delta_{k0}\delta_{j2}
\bigr)$
and
$C^A_{ijk}=\frac{1}{2}\omega_0\bigl(\delta_{i0}\delta_{j1}\delta_{k2}-\delta_{i0}\delta_{k1}\delta_{j2}
$
$-\delta_{i1}\delta_{j0}\delta_{k2}+\delta_{i1}\delta_{k0}\delta_{j2}\bigr)$.

Here, looking at the $k$-index tells us that the symmetric part
describes a small $x/y$-rotation as we move in $+z$-direction, but
also some $z/x$-shearing (``nonzero $z$-coordinate induces a change to
$x$-coordinate'') as we move along $+y$ as well as $z/y$-shearing as
we move along $+x$. Such ``shearing'' operations violate metric
compatibility -- if we parallel transport a $z$-directed and a
$y$-directed unit vector along the $+x$-direction, the angle between
them does not stay at 90 degrees. So, the symmetric part of the
connection coefficients $C^S$ violates metric compatibility,
but still gives us the same helical geodesic as the original
connection coefficients if we start with a vector such as
$\vec v_0=(10, 0, 1)$. While the $z$-component of that vector
only induces a rotation half as fast as the original rotation,
the ``shearing the $z$-component into the $y$-component'' that
gets multiplied with the $x$-velocity contributes the other half.

The antisymmetric part $C^A$ is called ``torsion''. It manifests as
follows: If we have a smooth scalar (number-valued) function $f$ on
some flat space with a metric on which we introduced Cartesian
coordinates, we can obtain the 2nd derivatives of $f$,
$\frac{\partial^2 f}{\partial x_i\,\partial x_j}$ at point
$\mathcal X$ with coordinates $X_i$ by evaluating $f$ at four
points and taking a limit to find the
rate-of-change-of-the-rate-of-change: if smoothness is guaranteed,
we have:
\begin{equation}
  \frac{\partial^2 f}{\partial x_i\,\partial x_j}=\lim_{\varepsilon\to 0}
  \Bigl(\left(f({\mathcal X}_D)-f({\mathcal X}_C\right)-
        \left(f({\mathcal X}_B)-f({\mathcal X}\right)
    \Bigr)/\varepsilon^2,
\end{equation}
where
${\mathcal X}_B$ is obtained by increasing the $i$-th coordinate by $\varepsilon$,
${\mathcal X}_C$ is obtained by increasing the $j$-th coordinate by $\varepsilon$,
and
${\mathcal X}_D$ is obtained by increasing the $i$-th and $j-th$
coordinate\footnote{If $i=j$, we perform an in total twice as large increase.}
by $\varepsilon$.

In curved space, there may well be an issue with how we obtain the
point ${\mathcal X}_D$: Conceptually, we pick two basis vectors
$\vec b^{(I)}, \vec b^{(II)}$ and find points $B, C$ by taking $\varepsilon$-steps
along these basis vectors. But how do we find point $D$? This seems to
ask for taking a $\varepsilon$-step from either $B$ or $C$ along the
direction of the vector that ``somehow'' corresponds to the other
basis vector, but at this displaced point. The most natural idea is to
obtain that direction by parallel-transporting the other basis vector
along the step we took, and then take a step along that other direction.
In general, there is no reason for the coordinate-correction of vector
$\vec b^{(I)}$ when parallel-transporting along vector $\vec b^{(II)}$ to equal
the coordinate-correction of vector $\vec b^{(II)}$ when parallel-transporting
along vector $\vec b^{(I)}$. In fact, if we take $\varepsilon$ steps, we expect
to get two points $D_1$ and $D_2$ whose coordinates differ to
second order in $\varepsilon$.

If we consider the point $D_0$ that we reach by not applying any
parallel transport correction when transporting displacement-vectors,
whose coordinates are $D_{0,i}=X_i+\varepsilon b^{(I)}_i+\varepsilon b^{(II)}_i$,
the coordinates of $D_1$ have an extra term that comes from taking the second
step not along $\varepsilon b^{(II)}_i$, but along the parallel-transported vector.
So, the coordinates of $D_1$ are $D_{1,i}=D_{0,i}-\varepsilon^2\,\sum_{j,k} C_{ijk}b^{(II)}_jb^{(I)}_k$,
while the coordinates of $D_2$ are $D_{2,i}=D_{0,i}-\varepsilon^2\,\sum_{j,k} C_{ijk}b^{(I)}_jb^{(II)}_k$.

The difference between the 2nd derivatives computed using $D_1$,
respectively $D_2$ equals
$\lim_{\varepsilon\to 0}\frac{1}{\varepsilon^2}\left(f(D_{1,i})-f(D_{2,i})\right)$.

Noting that the $D_{1,i},\,D_{2,i}$-coordinates have
$\propto\varepsilon^2$-corrections and substituting
$\mu:=\varepsilon^2$, we find that this difference is $-2\sum_{i,j,k}
C^A_{ijk}\frac{\partial f}{\partial x_i}b^{(I)}_jb^{(II)}_k$.

The term $2\,C^A_{ijk}$ is called ``torsion''. According to our
current (still limited) understanding of quantum gravity, energy (this
includes matter) is a source of curvature, while torsion arises from
nonzero intrinsic quantum spin (such as for the spin-1/2 Dirac field
of an electron). Curvature can propagate and influence physics away
from the sources, since the governing equations involve derivatives --
this is more basic than the existence of gravitational waves:
space-time around a star is curved also in regions where there is no
matter. In contrast, torsion is a purely local phenomenon that can
only exist where there is such matter with intrinsic spin: the
governing equations only involve physics at the point currently in
focus and no derivatives. In other words, torsion will not affect
geodesics in a ``no matter present'' vacuum region of curved space
time such as the black holes we are concerned with in this work.

Establishing the mathematical machinery that clarifies in which sense
and subject to what assumptions the aforementioned claims about
torsion hold would go well beyond the aspirations of this work. Here,
we only want to use the result from deeper theory that torsion is zero
in a vacuum, which implies that the connection coefficients are
symmetric under exchange of the final two indices. In that case, the
connection coefficients are uniquely defined if we postulate metric
compatibility\footnote{This is the fundamental theorem of
  (pseudo)-Riemannian Geometry.}. The corresponding connection is
known as the Levi-Civita connection. The connection coefficients are
then known as the Christoffel symbols (here: of the 2nd kind).

Let us start at a point (or, for space-time, time-and-position
`event') $\mathcal P$ with three local vectors
$\vec a^{\mathcal P}, \vec b^{\mathcal P}, \vec c^{\mathcal P}$.
If we consider parallel-transporting $\vec a^{\mathcal P}, \vec b^{\mathcal P}$
along some curve with momentary tangent $\vec c^{\mathcal P}$,
the rate-of-change of the scalar product as we do so must be zero.

More concretely, the rate-of-change of the scalar product for
parallel-transporting by $\varepsilon\cdot\vec c^{\mathcal P}$, where
we imagine $\varepsilon$ to gradually increase from zero to some small
quantity, must be zero at $\varepsilon=0$. Structurally, the
corresponding equation is of the form
\begin{equation}
  \begin{array}{lcl}
    0&=&\left.\frac{\partial}{\partial\varepsilon}\right|_{\text{at}\,\varepsilon=0}\\
    &&\left(M_{ij}^{\text{at} {\mathcal P}}+\varepsilon\cdot\sum_k c_k\cdot
    \left\{
    \begin{array}{l}
      \text{\scriptsize rate of change of}\,M_{ij}\\
      \text{\scriptsize along coordinate}\;k
    \end{array}
    \right\}
    \right)\times\\
     &&\left(a_i-\varepsilon\sum_{p, q}C_{ipq}a_pc_q\right)\times\\
     &&\left(b_j-\varepsilon\sum_{r, s}C_{jrs}b_rc_s\right).
  \end{array}
\end{equation}

The rate of change with respect to $\varepsilon$ at $\varepsilon=0$ is
neither sensitive to any higher order terms in the rate-of-change of
the entries of $M$ with coordinates nor to any terms that arise if one
of the $\varepsilon$-terms above multiplies another such
$\varepsilon$-term in the product above. After applying the
distributive law, the ``$\varepsilon$-derivative at $\varepsilon=0$''
picks out the coefficient of $\varepsilon^1$, which contains
contributions from the coordinate-dependency of the metric tensor's
entries as well as the parallel transport corrections. After some
simple index gymnastics, we get this constraint that ties the
connection coefficients to the coordinate derivatives of the metric:
\begin{equation}
  \begin{array}{lcl}
    0&=&\sum_kc_k\cdot\Bigl(\sum_{i,j}a_ib_j\,\frac{\partial M_{ij}}{\partial x_k}\\
    &&-\sum_{i,j,p} M_{ij}C_{ipk}a_pc_kb_j\\
    &&-\sum_{i,j,p} M_{ij}C_{jpk}a_ic_kb_p\Bigr).
  \end{array}
\end{equation}

Since this must hold for every $c_k$, so especially when using each
coordinate basis vector as $\vec c^{\mathcal P}$, we get the following
collection of relations, one for every $k$, where we use the common
short hand $M_{ij,k}$ for $\partial M_{ij}/\partial x_k$ and adopt a
``indices appearing twice in a product of multi-index objects automatically
are summed over'' approach:
\begin{equation}
M_{ij}C_{ipk}a_pb_j + M_{ij}C_{jpk}a_ib_p = M_{ij,k}a_i b_j.
\end{equation}

Relabeling indices so that the $a$-index is always $i$ and $b$-index
is always $j$, using $M_{ij}=M_{ji}$ and employing the same argument
as above for eliminating $c_k$, we get:
\begin{equation}
M_{jp}C_{pik} + M_{ip}C_{pjk} = M_{ij,k}.
\end{equation}

Before we determine $C$, let us first focus on finding an equation for
$M\cdot C$. The right hand side is easy to obtain for any $i,j,k$ if
we know the metric tensor's coordinate dependency -- and it does not
matter if we are given this dependency in symbolic or in algorithmic
form. The second summand on the left hand side arises from the scalar
product of the $i$-th basis vector (which we ultimately here used as
$\vec a$) with the parallel-transport correction of the $j$-th basis
vector (which we used as $\vec b$) when parallel-transporting along
the $k$-th basis vector (which we used as $\vec c$). Right as it is
here, we cannot express this contribution in terms of unknown
quantities since it gets added to another such term that involves the
connection coefficients which involves the connection coefficients
correcting the other vector. Is there anything else we can say about
$M_{ip}C_{pjk}$ -- perhaps some other relation that gives rise to this
quantity? At this point, we can invoke the ``torsion is zero''
symmetry of the connection in the last two indices, $C_{pjk}=C_{pkj}$:
We can swap the roles of the 2nd parallel-transported vector $\vec b$
and the transport direction $\vec c$, which here amounts to swapping
$j$ and $k$:
\begin{equation}
M_{kp}C_{pij} + M_{ip}C_{pkj} = M_{ik,j}.
\end{equation}

Our construction was manifestly symmetric under swapping the roles of
$\vec a^{\mathcal P}$ and $\vec b^{\mathcal P}$, so if we swap
$\vec b^{\mathcal P}\leftrightarrow \vec c^{\mathcal P}$ as we just did,
we may just as well have swapped
$\vec a^{\mathcal P}\leftrightarrow \vec c^{\mathcal P}$.
We can still do this by swapping $i\leftrightarrow j$ above
(and using $C_{pqr}=C_{prq}$):
\begin{equation}
  M_{kp}C_{pij} + M_{jp}C_{pki} = M_{jk,i}.
\end{equation}

Let us simplify this further by introducing $Z(i;j,k):= M_{ip}C_{pjk}$
as the contribution to the momentary rate-of-change of the scalar
product between the $i$-th and $j$-th basis vector that we get when
parallel transporting both along a curve with momentary tangent the
$k$-th basis vector that is due to the coordinate correction of the
$j$-th basis vector. Due to $j\leftrightarrow k$ symmetry for
torsion-free connections, for any given set of three (not necessarily
different) basis vectors, we can form at most $3!/2=3$ Z's from these,
and the equations above each give us information about the sum of a
pair of these. This structure arises since ``parallel transport of a
scalar product'' involves two vector-corrections together, while we
are interested in determining how to correct a single vector. In terms
of the $Z$'s, we have:

\begin{equation}
  \begin{array}{lcl}
    Z(j;ik) + Z(i;jk) & = & M_{ij,k}\\
    Z(k;ij) + Z(i;jk) & = & M_{ik,j}\\
    Z(k;ij) + Z(j;ik) & = & M_{jk,i}
  \end{array}
\end{equation}

Structurally, we are in the situation of a customer of a quirky vendor
that only sells items in their inventory in non-matching pairs
(like pairs of socks of different colors). For any set of three colors,
such as red, green, blue ($i,j,k)$, if we wanted to infer the price of a single
green sock, we can find this by buying a green/red pair plus a green/blue pair,
then subtracting the price of a red/blue pair and dividing by two.
Correspondingly, adding the first and second equations (which both have $Z(i;jk)$)
and subtracting the third gets us:
\begin{equation}
  \begin{array}{l}
    Z(j;ik) + Z(k;ij) + 2Z(i;jk)  - Z(k;ij) - Z(j;ik) =\\
    M_{ij,k} + M_{ik,j}-M_{jk,i},
  \end{array}
\end{equation}
and using the definition of $Z$, we find
\begin{equation}
  M_{ip}C_{pjk} = \frac{1}{2}\left(M_{ij,k}+M_{ik,j}-M_{jk,i}\right).
\end{equation}

Finally, we can remove the factor $M_{ip}$ on the left hand side that
binds the $p$-index on the $C_{pjk}$ and leaves us with an $i$-index
by transforming an equation with index structure $X_{ijk} = Y_{ijk}$
into $R_{mi}X_{ijk}=R_{mi}Y_{ijk}$ here with $R_{mi} = M^{(-1)}_{mi}$,
i.e. $R_{mi}M_{ip}=\delta_{mp}$. Doing this and un-doing all the
abbreviations we introduced gets us to, after relabeling indices:
\begin{equation}
  C_{ijk} = \frac{1}{2}M^{(-1)}_{ip}\left(\frac{\partial M_{pj}}{\partial x_k} + \frac{\partial M_{pk}}{\partial x_j} - \frac{\partial M_{jk}}{\partial x_p}\right).
\end{equation}

\section{An Example for Calculating Christoffel Symbols}\label{app:ChristoffelExample}

Now that we know the Christoffel symbols are derived directly from the
metric, let us work through a concrete example.  We will use the
two-dimensional polar coordinate system, which we introduced earlier.
Its geometry is defined by the line element, which gives the squared
distance between two infinitesimally close points:
\begin{equation}
  (ds)^2 = (dr)^2 + r^2\,(d\theta)^2.
\end{equation}

From the coefficients of this line element, we can write down the metric
tensor:
\begin{equation}
  M = \begin{pmatrix} 1 & 0 \\ 0 & r^2 \end{pmatrix},
  \qquad
  M^{-1} = \begin{pmatrix} 1 & 0 \\ 0 & 1/r^2 \end{pmatrix}.
\end{equation}

The general formula for the Christoffel symbols, as derived in
Appendix~\ref{app:Christoffel}, is:
\begin{equation}
  \Gamma^i{}_{k\ell}
    = \frac{1}{2} \sum_m (M^{-1})^{im}
      \left(
        \frac{\partial M_{mk}}{\partial x^\ell}
        + \frac{\partial M_{m\ell}}{\partial x^k}
        - \frac{\partial M_{k\ell}}{\partial x^m}
      \right).
\end{equation}

For polar coordinates, our coordinate indices are $0=r$ and $1=\theta$.
The only non-zero partial derivative of a metric component is:
\begin{equation}
  \frac{\partial M_{11}}{\partial r} = 2r.
\end{equation}

All other derivatives are zero.  Now we can use the formula to calculate
the Christoffel symbols.  Many of them will be zero.  We will only
calculate the ones we know will be non-zero.

{\bf Calculating $\Gamma^0{}_{11}$}.
Here, we have $i=0$, $k=1$, and $\ell=1$.  The sum is over the
index~$m$.  Since the inverse metric $M^{-1}$ is diagonal, the only
non-zero term in the sum is when $m=0$:
\begin{equation}
  \begin{array}{lcl}
  \Gamma^0{}_{11}
    & = & \tfrac{1}{2}\,(M^{-1})^{00}
      \Bigl(
        \underbrace{\tfrac{\partial M_{01}}{\partial x^1}}_{=\,0}
        + \underbrace{\tfrac{\partial M_{01}}{\partial x^1}}_{=\,0}
        - \tfrac{\partial M_{11}}{\partial x^0}
      \Bigr)\\
    & = & \tfrac{1}{2}(1)(0 + 0 - 2r) = -r.
    \end{array}
\end{equation}

So, $\Gamma^0{}_{11} = -r$.  This non-zero value tells us that when
parallel-transporting a vector in the angular direction, the vector's
radial component will change.  Specifically, a vector with only an
angular component (like the basis vector~$e_\theta$) must acquire a
radial component in order to stay ``parallel transported.''  This makes
intuitive sense because as one moves around a circle, one's local notion
of ``straight ahead'' is constantly bending.

{\bf Calculating $\Gamma^1{}_{01}$ and $\Gamma^1{}_{10}$}.
Here, we have $i=1$, $k=0$, and $\ell=1$.  Due to the symmetry of the
connection, we know $\Gamma^1{}_{01} = \Gamma^1{}_{10}$.  The only non-zero
term is when $m=1$:
\begin{equation}
  \Gamma^1{}_{01}
    = \tfrac{1}{2}\,(M^{-1})^{11}
      \Bigl(
        \underbrace{\tfrac{\partial M_{10}}{\partial x^1}}_{=\,0}
        + \tfrac{\partial M_{11}}{\partial x^0}
        - \underbrace{\tfrac{\partial M_{01}}{\partial x^1}}_{=\,0}
      \Bigr)
    = \frac{1}{2r^2}(2r)
    = \frac{1}{r}.
\end{equation}

So, $\Gamma^1{}_{01} = \Gamma^1{}_{10} = \frac{1}{r}$.  This non-zero
value indicates that as we change our radial position, the angular basis
vector~$\vec{e}_\theta$ must ``twist'' in order to remain ``parallel.''

For the polar coordinate system, the only non-zero Christoffel symbols
are:
\begin{equation}
  \Gamma^0{}_{11} = -r,
  \qquad
  \Gamma^1{}_{01} = \Gamma^1{}_{10} = \frac{1}{r}.
\end{equation}
  
This example shows how the (local) Christoffel symbols, which define
parallel transport, are fully determined by the (local) metric tensor
and its (local) spatial derivatives.

The Christoffel symbols carry three coordinate-indices; at any given
point, we could perform a coordinate transformation from the
coordinate system that uses the $(r, \theta)$ coordinate-lines to the
coordinate system that uses the $(x, y)$ coordinate-lines. In the
Cartesian coordinate system, we have no parallel transport correction,
so the Christoffel symbols are all zero. Yet, applying a coordinate
transformation to nonzero Christoffel symbols (like the ones for polar
coordinates) cannot produce zero Christoffel symbols. In the General
Relativity literature, one frequently finds the claim that
``Christoffel symbols do not transform like a tensor'', but this is
mostly based on a misunderstanding: If we perform a coordinate
transformation to express the Christoffel symbols for polar
coordinates in terms of Cartesian basis vectors, they do not magically
become Christoffel symbols for Cartesian coordinates -- they still are
the Christoffel symbols for polar coordinates, only expressed in
a local basis that is appropriate for a different coordinate system\footnote{
We see the same motif in the joke about the wife complaining to her husband
that their neighbor kisses his wife goodbye every morning when he leaves
for work, while the addressee of said complaint never does this -- responds
the husband ``why would I -- I barely know her?''.}.

\section{On the Meaning of the Christoffel Symbols}\label{app:ChristoffelPhysics}

To make the concept of Christoffel symbols representing a ``fictitious
force'' even more concrete, let us consider the famous ``Einstein's
elevator'' thought experiment. Imagine we are in a sealed, ground-fixed
elevator on Earth, holding a ball. We drop the ball, and it falls to the
floor with an acceleration of $9.8\,\text{m/s}^2$. The force one feels
holding the ball is gravity. Now imagine we are in an identical elevator
far out in space, away from any gravitational sources. If we press a
button to accelerate the elevator ``upwards'' at $9.8\,\text{m/s}^2$,
the floor will push up, and if we release the ball, the floor will rush
up to meet it, making it appear as if the ball accelerated towards the
floor. In both scenarios, the internal experience is identical. This
extends to the passengers themselves: the passenger on Earth feels their
own weight pushing them down, while the passenger in space feels the
accelerating floor pushing up against their feet.

From the perspective of an observer outside the elevator in space (an
inertial frame), there is no force on the ball. The ball is simply
following a ``straight path'' (a geodesic). The apparent acceleration of
the ball is caused entirely by the fact that the observer inside the
elevator is in a non-inertial, accelerating coordinate system. The
Christoffel symbols are the mathematical objects that describe exactly
this phenomenon.

While directly applying our formula for $\Gamma$ from a metric is
tricky here (as Newtonian physics is not built on a spacetime metric
in the same way as GR), the result for the accelerating coordinate
system is intuitive. If the elevator's spatial elevation-coordinate is
$x$ and it accelerates upwards at $a$, the only significant non-zero
Christoffel symbol is $C_{100}=C_{`xtt'}=\Gamma^x{}_{tt}=a$. This term
represents the ``fictitious force.'' In the geodesic equation, this
component is the downward acceleration, inducing a coordinate
acceleration of $-a$ for a stationary object (one whose only velocity
is through time) to follow a ``straight'' path: Forward-propagating
the momentarily-at-rest object's purely future-directed ``worldline
trajectory'' tangent makes its forward-in-time-motion generate a
downward-pointing component of their relative-to-the-elevator
velocity-vector\footnote{For rotating coordinate systems, we would find the
Coriolis forces in entries such as $\Gamma^y{}_{tx}=\Gamma^y{}_{xt}$.}.

Gravity is not a force in the traditional sense, but a manifestation of
the curvature of spacetime itself, which appears to us as a
coordinate-dependent acceleration. This is why an astronaut in orbit is
``weightless''. They are in free-fall, following a geodesic in curved
spacetime. From their perspective, they feel no forces and no
acceleration, as they are on the straightest possible path.

\onecolumn

\section{The Complete Construction}\label{app:FullCode}

This appendix contains the complete, skeletal construction in under
380 lines total of code plus diligent documentation. The raw code
itself is under 180 lines.

\begin{lstlisting}[language=Python]
r"""Minimalistic self-contained JAX-based General Relativity Raytracer.

Many functions in this module take or return arguments that themselves
are functions with specific rule in General Relativity. In order to
eliminate repetitiveness in the documentation, these are described
upfront here, with individual functions referring to the
module-docstring where needed. The key definitions are:

  L2_func: Callable `f(coords_pos, coords_vec) -> len_sq` which
    computes the length-squared of a local (at position given by
    the `[dim]-jax.Array` `coords_pos`) vector with
    'spatial coordinate rate-of-change per curve-parameter rate-of-change'
    coordinates `coords_vec`, given as another `[dim]-jax.Array`.

  metric_func: Callable `g_func(coords_pos) -> coords_metric`
    that maps position-coordinates (given as a `float64 [dim]-jax.Array`)
    to metric-tensor coordinates (as a `float64 [dim, dim]-jax.Array`).

  christoffel_func: Callable `christoffel(coords_pos) -> christoffel_symbols`
    that maps position-coordinates  (given as a `[dim]-jax.Array`) to a
    `[dim, dim, dim]-jax.Array` - henceforth `Gamma` - containing the
    Christoffel symbols of the 2nd kind at that point. In conventional
    notation: `Gamma[m, n, p] = \Gamma^m_np`.

  ode_y_func: Callable `F(y) -> rate_of_change_of_y` for numerical
    ODE-integration. Here, `y` is a `jax.Array` such that
    `tuple(y.reshape(2, dim))` is the tuple `(coords_pos, coords_tangent)`
    containing position and tangent vector data.

The term 'basis vector' refers to a tangent vector on a coordinate-line curve.
"""

import pdb  # For: `python3 -i CODE.py` and `pdb.pm()` inspection on `raise ...`
import itertools, matplotlib.pyplot, numpy, os, warnings

import jax
from jax import numpy as jnp
jax.config.update("jax_enable_x64", True)


def metric_func_from_L2_func(L2_func):
  "Computes `metric_func` function from `L2_func` function."
  d2_L2_by_dxi_dxj_func = jax.hessian(L2_func, argnums=1)
  def metric_func(coords_pos):
    """Maps X_i -> M_ij, with X=position, M=metric tensor."""
    return 0.5 * d2_L2_by_dxi_dxj_func(coords_pos, jnp.zeros_like(coords_pos))
  return metric_func


def christoffel_func_from_metric_func(metric_func):
  """Maps 'metric tensor' function to 'Christoffel symbols' function.

  Args:
    metric_func: See `metric_func` in module docstring.
  Returns:
    See `christoffel_func` in module docstring.
  """
  # {coords} -> d {metric} / d {coord}
  d_metric_d_coords_func = jax.jacobian(metric_func)
  def christoffel_func(coords_pos):
    """Maps X_i -> C_ijk, with X=position, C=Christoffel symbols of 2nd kind."""
    # These are the g_ij,k (or M_ij,k):
    d_metric_d_coords_pos = d_metric_d_coords_func(coords_pos)
    # We use jnp.einsum() rather than Array.transpose() and also
    # jnp.linalg.inv() rather than jnp.linalg.solve() for better
    # alignment with the text.
    return 0.5 * jnp.einsum('im,mkl->ikl',
                            jnp.linalg.inv(metric_func(coords_pos)),
                            d_metric_d_coords_pos
                            + jnp.einsum('mlk->mkl', d_metric_d_coords_pos)
                            - jnp.einsum('klm->mkl', d_metric_d_coords_pos))
  return christoffel_func


def geodesic_dy_dt_func_from_christoffel_func(christoffel_func):
  "Computes `ode_y_func` function from `christoffel_func` function."
  def dy_dt_func(motion_state_y):
    """Maps Y_i -> {rate-of-change of Y_i}."""
    c_pos, c_tangent = jnp.split(motion_state_y, 2)
    rate_tangent = -jnp.einsum('ijk,j,k->i', christoffel_func(c_pos),
                               c_tangent, c_tangent)
    return jnp.concatenate([c_tangent, rate_tangent], axis=0)
  return dy_dt_func


def geodesic_dy_dt_func_from_metric_func(metric_func):
  "Computes `ode_y_func` function from `metric_func` function."
  christoffel_func = christoffel_func_from_metric_func(metric_func)
  return geodesic_dy_dt_func_from_christoffel_func(christoffel_func)    


# This actually generalizes to a wider class of functions than `ode_y_func`.
def rk4_estimate(f, y0, ds):
  """Estimates average `dy/dt` over step `ds` via Runge-Kutta RK4.

  Args:
    f: See `ode_y_func` in module docstring.
    y0: `jax.Array` representing the current motion-state.
    ds: step-size parameter interval. We try to estimate the average `dy/dt`
        over a step of this size.

  Returns:
    RK4 estimate of average `dy/dt` when using `f` to propagate motion-state
    over interval `ds`.
  """
  k1 = f(y0)
  k2 = f(y0 + (0.5 * ds) * k1)
  k3 = f(y0 + (0.5 * ds) * k2)
  k4 = f(y0 + ds * k3)
  return (k1 + 2.0 * k2 + 2.0 * k3 + k4) / 6.0


def ode_solve(estimate_func, f, y0, s_samples):
  """Numerically integrates an ordinary differential equation.

  Args:
    estimate_func: mean-rate-of-change estimating function.
      Must have the same signature as `rk4_estimate`.
    f: See `ode_y_func` in module docstring.
    y0: `jax.Array` representing the current motion-state.
    s_samples: 1-axis `jax.Array` with coordinate-parameter sample-values.
      ODE-integration attributes motion-state `y0` to parameter-value
      `s_samples[0]` and then proceeds in steps of
      `s_samples[n+1] - s_samples[n]`.
  Returns:
    `[num_steps, *y0.shape()]-jax.Array` `result` with collected
    motion-state-at-end-of-every-step data. Here, `result[k]` corresponds to
    parameter-value `s_samples[k+1]`, and `num_steps + 1 == s_samples.size`.
  """
  def inner_func(y_now, s_step):
    rate = estimate_func(f, y_now, s_step)
    y_next = y_now + s_step * rate
    return y_next, y_next
  y_final, ys_collected = jax.lax.scan(inner_func, y0, jnp.diff(s_samples))
  return ys_collected


def riemann_func_from_christoffel_func(christoffel_func):
  """Computes Riemannian curvature function from Christoffel function.

  Args:
    christoffel_func: See `christoffel_func` in module docstring.

  Returns:
    Callable `R(coords_pos) -> curvature_tensor` that maps position to
    a `[dim, dim, dim, dim]-jax.Array` - henceforth `Riemann` - containing
    the Riemann curvature tensor. In conventional GR notation:
    `Riemann[m, n, p, q] == R^m_npq`.
  """
  jacobian_christoffel_func = jax.jacobian(christoffel_func)
  def riemann_func(coords_pos):
    """Maps X_i -> R^m_npq, where X=position, R=Riemann curvature tensor."""
    christoffel = christoffel_func(coords_pos)
    d_christoffel = jacobian_christoffel_func(coords_pos)
    # We use jnp.einsum() throughout, also for simple transposition.
    aux = (jnp.einsum('ijlk->ijkl', d_christoffel) +
           jnp.einsum('ikm,mjl->ijkl', christoffel, christoffel))
    # The whole expression is aux_ijkl - aux_ijlk.
    return aux - jnp.einsum('ijkl->ijlk', aux)
  return riemann_func


def einstein_func_from_L2_func(L2_func):
  """Computes Einstein Tensor function from Line Element function.

  Args:
    L2_func: See `L2_func` in module docstring.

  Returns:
    Callable `G(coords_pos) -> einstein_tensor` that maps position to
    a `[dim, dim]-jax.Array` - henceforth `Einstein` - containing
    the Einstein curvature tensor. In conventional GR notation:
    `Einstein[m, n] == G_mn`.
  """
  metric_func = metric_func_from_L2_func(L2_func)
  christoffel_func = christoffel_func_from_metric_func(metric_func)
  riemann_func = riemann_func_from_christoffel_func(christoffel_func)
  def einstein_func(coords):
    """Maps X_i -> G^mn, where X=position, G=Einstein tensor."""
    metric = metric_func(coords)
    riemann = riemann_func(coords)
    ricci = jnp.einsum('mimj->ij', riemann)
    return (ricci - 0.5 * jnp.einsum('ij,ij->', ricci, jnp.linalg.inv(metric))
            * metric)
  return einstein_func


def schwarzschild_L2_func(txyz, d_txyz, r_schwarzschild=1.0):
  """Computes the local-vector length-squared for Schwarzschild geometry.

  Schwarzschild geometry is given by:
  ```
  ds**2 = -F(r)*dt**2 + dr**2/F(r) +
           + r**2 d_theta**2 + r**2 * sin(theta)**2 d_phi**2
  ```

  ...where here, we have `F(r) = 1/(1-r_schwarzschild/r)`, `r*cos(theta)=z`,
  `r*sin(theta)*cos(phi)=x`, `r*sin(theta)*sin(phi)=y`, i.e. `theta`
  is the angle from the 3d "north" axis.
  
  Args:
    txyz: `[4]-jax.Array` position coordinate-vector `(t, x, y, z)`.
    d_txyz: `[4]-jax.Array`, tangent vector (w.r.t. local coordinate
      rate-of-change vector space basis).
    r_schwarzschild: The "Schwarzschild radius".

  Returns:
    The value of the "space-time" (local Minkowski) pseudo-Euclidean
    scalar product of the vector with coordinates `d_txyz` with itself,
    or `jax.numpy.nan` if `r <= r_schwarzschild`.
  """
  xyz = txyz[1:]
  r = jnp.linalg.norm(xyz)
  e_r = xyz / r
  part_e_r = (d_txyz[1:] @ e_r) * e_r
  part_perpendicular = d_txyz[1:] - part_e_r
  ds2_perpendicular = jnp.square(part_perpendicular).sum()
  s_factor = jnp.where(r > r_schwarzschild, 1 - r_schwarzschild / r, jnp.nan)
  ds2_radial = jnp.square(part_e_r).sum() / s_factor
  ds2_time_abs = jnp.square(d_txyz[0]) * s_factor
  return ds2_radial + ds2_perpendicular - ds2_time_abs


# Validation of the construction - can be removed:
schwarzschild_einstein_func = einstein_func_from_L2_func(schwarzschild_L2_func)
jax_schwarzschild_G = jax.jit(schwarzschild_einstein_func)
einstein1 = jax_schwarzschild_G(jnp.array([5.0, 2.0, 3.0, 1.0]))
print('=== Einstein tensor at an arbitrary point (should be zero) ===')
print(einstein1.round(12))

#### Scaffolding

def get_orthonormal_frame(metric_func, txyz, frame_raw):
  """Obtains an orthonormal coordinate frame from "future" / "forward" / "up".

  Here, `coords_dir_*` vectors are at point `txyz` and w.r.t. `txyz`
  coordinate-lines tangent vector basis.
  
  Args:
    metric_func: See `metric_func` in module docstring.
    txyz: Space-time coordinates of the camera.
    frame_raw: [4, 4]-ArrayLike `r` such that `r[:, 0]` will be proportional
      to the "forward-in-time" unit-vector of the resulting orthonormal frame,
      the projection of `r[:, 1]` to the subspace orthogonal to `r[:, 0]`
      will be proportional to the resulting orthonormal frame's "x" ("forward")
      direction, and the projection of `r[:, 3]` to the subspace orthogonal to
      the "t" and "x" frame-directions will be proportional to "z".
  """
  g = metric_func(txyz)
  sprod = lambda v1, v2: numpy.einsum('ij,i,j->', g, v1, v2)
  frame_raw_txzy = numpy.asarray(frame_raw,
                                 dtype=numpy.float64)[:, (0, 1, 3, 2)]
  basis_vecs = []
  for n, b in enumerate(frame_raw_txzy.T):
    b_now = b
    for b_earlier in basis_vecs:
      b_now = b_now - b_earlier * (sprod(b_now, b_earlier) /
                                   sprod(b_earlier, b_earlier))
    basis_vecs.append(b_now / abs(sprod(b_now, b_now))**.5)
  return numpy.stack(basis_vecs, axis=1)[:, (0, 1, 3, 2)]


def _get_ray_eye_ends(txyz_end, frame, dyzs, dx=1, dt=1, c=1):
  # Find trajectory end-state-vectors that reach us going
  # in the 3d direction (w.r.t. local frame) `-(-dx, y, z)`.
  # (We ODE-integrate rays "backwards in time").
  dyzs_flat = dyzs.reshape(-1, 2)  # indexed `[num_pixel, yz_coord]`.
  dxyzs_flat = jnp.pad(-dyzs_flat, ((0, 0), (1, 0)), constant_values=dx)
  # We are here working with spatial coordinates in a locally-orthonormal frame.
  normed_dxyzs_flat = dxyzs_flat / jnp.linalg.norm(
    dxyzs_flat, axis=-1, keepdims=True)
  dtxyzs_flat = jnp.pad(c * dt * normed_dxyzs_flat,
                        ((0, 0), (1, 0)), constant_values=dt)
  ode_states_flat = (
    # eye-point (txyz) coordinates, broadcastable across pixel-index,
    # padded to accomodate 4d tangent-coordinates.
    jnp.pad(txyz_end[jnp.newaxis, :], ((0, 0), (0, 4)))
    # per-pixel 4d tangent coordinates.
    + jnp.pad(jnp.einsum('cL,nL->nc', frame, dtxyzs_flat), ((0, 0), (4, 0))))
  return ode_states_flat.reshape(*dyzs.shape[:-1], 4 + 4)


#### Rendering

if 'DEMO-RENDER':
  CHECKER_SCALE = 8
  NUM_PATCHES_XY, PATCH_NUM_PIXELS_XY = 8, 72
  PIXELS_XY = PATCH_NUM_PIXELS_XY * NUM_PATCHES_XY
  z_displacement = 0.1  # How far to move off of the z=0 plane.
  color_ud, color_du = numpy.array([1, 1, 0, 1, 0.5, 0]).reshape(2, 1, 1, 3)  
  ode_steps = jnp.linspace(0, 100.0, 401, dtype=jnp.float64)
  jax_get_ray_eye_ends = jax.jit(_get_ray_eye_ends)
  r_disc_min, r_disc_max = 4, 8
  color_ud, color_du = numpy.array([1, 1, 0, 1, 0.5, 0]).reshape(2, 1, 1, 3)
  # Black Hole ("BH") Geometry:
  bh_metric_func = metric_func_from_L2_func(schwarzschild_L2_func)
  bh_dy_dt_func = geodesic_dy_dt_func_from_metric_func(bh_metric_func)
  def ray_trace(y0):
    return ode_solve(rk4_estimate, bh_dy_dt_func, y0, ode_steps)
  batch_ray_trace_flat = jax.vmap(ray_trace)
  @jax.jit
  def batch_ray_trace(y0s):
    return (batch_ray_trace_flat(y0s.reshape(-1, y0s.shape[-1]))
            .swapaxes(0, 1).reshape(-1, *y0s.shape))
  # Camera:
  eye_coords = 20.0 * jnp.array([0.0, 1.0, 0.0, z_displacement],
                                dtype=jnp.float64)
  camera_frame = get_orthonormal_frame(
    bh_metric_func, eye_coords,
    [(1,  0, 0, 0),       # t
     -eye_coords,         # forward (towards black hole)
     (0, 0, -1, 0),       # "y" - negative to make right-handed xyz.
     (0,  0, 0, 1)])      # "up"
  print('DDD Camera Frame\n', camera_frame.round(6).tolist())
  coord_steps = jnp.linspace(-0.75, 0.75, PIXELS_XY)
  # Target data:
  full_image = numpy.zeros((PIXELS_XY, PIXELS_XY, 3))
  image_blocks = full_image.reshape(
    *(NUM_PATCHES_XY, PATCH_NUM_PIXELS_XY)*2, 3).transpose(0, 2, 1, 3, 4)
  xy_coords = jnp.stack(
    jnp.meshgrid(coord_steps, coord_steps, indexing='xy'), axis=-1)
  xy_patches = xy_coords.reshape(
    # Splitting both i,j image-indices into "slow" block and fast
    # "pixel in block" indices - and pullung block-indices to the front:
    *(NUM_PATCHES_XY, PATCH_NUM_PIXELS_XY)*2, 2).transpose(0, 2, 1, 3, 4)
  for n_block_row, n_block_col in itertools.product(
      range(NUM_PATCHES_XY), repeat=2):
    xy_patch = xy_patches[n_block_row, n_block_col]
    ray_eye_ends = jax_get_ray_eye_ends(eye_coords, camera_frame, xy_patch)
    ray_traced_patch = numpy.array(batch_ray_trace(ray_eye_ends))
    # If the "far" end of the ray is not far behind the hole,
    # we color the current pixel white.
    ray_far_ends_x = ray_traced_patch[-1, ..., 1]
    ray_far_ends_dxyz = ray_traced_patch[-1, ..., -3:]
    y_cell_index, z_cell_index = (numpy.round(
      CHECKER_SCALE * ray_far_ends_dxyz[..., d] / ray_far_ends_dxyz[..., 0])
       for d in (1, 2))
    checkerboard_index = (y_cell_index + z_cell_index) % 2
    # Here, we expect to see three values: 0, 1, NaN.
    image_patch = numpy.where(
      # De-noising the interior uses some intermediate numpy trickery.
      numpy.isfinite(numpy.lib.stride_tricks.sliding_window_view(
        numpy.pad(ray_far_ends_x, ((1, 1), (1, 1))), (3, 3)
      )).all(axis=(-1, -2)),
      # Case: we did not hit the hole.
      numpy.where(
        (ray_far_ends_dxyz[..., 0] < 0) & (ray_far_ends_x < -10),              
        # We did pass the hole and see the checkerboard sky behind it.
        (0.25 + 0.5 * checkerboard_index),
        1.0),  # Case: We did not pass the hole - "white on the map".
      0.0)  # Case: We did hit the hole - "black on the map".
    image_blocks[n_block_row, n_block_col, ...] = image_patch[..., jnp.newaxis]
    # Adding an accretion disc:
    z_flat = ray_traced_patch[:, :, :, 3].reshape(ray_traced_patch.shape[0], -1)
    r_flat = numpy.linalg.norm(
      ray_traced_patch[:-1, :, :, 1:4], axis=-1).reshape(z_flat[:-1, :].shape)
    r_in_range = (r_disc_min < r_flat) & (r_flat < r_disc_max)
    z_transition = numpy.diff(numpy.sign(z_flat), axis=0)
    first_transition_index = numpy.argmax(
      r_in_range & z_transition.astype(bool), axis=0)
    z_transition_type = z_transition[
      first_transition_index,
      numpy.arange(first_transition_index.size)].reshape(*image_patch.shape, 1)
    image_blocks[n_block_row, n_block_col, :] = numpy.where(
      z_transition_type == 0,
      image_blocks[n_block_row, n_block_col, :],  # No change here.
      numpy.where(z_transition_type > 0, color_du, color_ud))
    print(f" - Rendered patch ({n_block_row}, {n_block_col})")
  matplotlib.pyplot.imshow(full_image)
  matplotlib.pyplot.show()
\end{lstlisting}

\twocolumn

\end{document}